\documentclass[12pt]{iopart}

\usepackage{here}
\usepackage{graphicx}
\usepackage{float}
\usepackage{epsfig}
\usepackage{cite}
\usepackage{url}
\usepackage{helvet}
\usepackage[dvips]{color}

\newcommand{\beq}{\begin{equation}}
\newcommand{\eeq}{\end{equation}}
\newcommand{\lpa}{\left(}
\newcommand{\rpa}{\right)}
\newcommand{\lpb}{\left[}
\newcommand{\rpb}{\right]}

\newcommand{\kpar}{k_{\|}}
\newcommand{\npar}{n_{\|}}

\newcommand{\eps}{\epsilon}
\newcommand{\epsL}{\epsilon_{\rm L}}
\newcommand{\epsR}{\epsilon_{\rm R}}
\newcommand{\epsS}{\epsilon_{\rm S}}

\newcommand{\Xhe}{X[{\rm ^{3}He}]}

\newcommand{\Xcrhe}{X_{\rm crit}[{\rm ^{3}He}]}

\newcommand{\Hy}{\rm H}
\newcommand{\He}{^{3}{\rm He}}

\newcommand{\Be}{\rm Be}
\newcommand{\Wo}{\rm W}

\newcommand{\W}[1]{{\rm W}^{{#1}+}}
\newcommand{\C}{{\rm C}^{6+}}
\newcommand{\HeH}{\textrm{($^{3}$He)--H}}

\newcommand{\HeD}{\textrm{($^{3}$He)--D}}
\newcommand{\HDe}{\textrm{(H)--D}}

\newcommand{\ntor}{n_{\rm tor}}

\newcommand{\om}{\omega}
\newcommand{\omn}{\tilde{\omega}}

\newcommand{\omp}[1]{\omega_{p{#1}}}
\newcommand{\omc}[1]{\omega_{c{#1}}}

\newcommand{\ompi}{\omega_{pi}}
\newcommand{\omci}{\omega_{ci}}

\newcommand{\ompe}{\omega_{pe}}
\newcommand{\omce}{\omega_{ce}}

\newcommand{\omnl}{{\tilde{\omega}}_{\rm L}}
\newcommand{\omns}{{\tilde{\omega}}_{\rm S}}

\newcommand{\ffraci}{\frac{\om}{\sqrt{2} \kpar v_{ti}}}
\newcommand{\ffrac}[1]{\frac{\om}{\sqrt{2} \kpar v_{t{#1}}}}

\newcommand{\zzetai}[1]{Z(\xi_{{#1}i})}

\newcommand{\zn}{\mathcal{Z}}

\newcommand{\ws}{\omega_{\rm S}}
\newcommand{\wns}{\tilde{\omega}_{\rm S0}}
\newcommand{\wl}{{\omega_{\rm L}}}
\newcommand{\wnl}{\tilde{\omega}_{\rm L0}}

\newcommand{\imp}{\rm imp}
\newcommand{\crit}{\rm crit}

\begin{document}

\title[Effect of impurities on the transition between minority ion and MC ICRH heating]
{Effect of impurities on the transition 
between minority ion and mode conversion ICRH heating
in~$\bf{(^{3}He)}$--$\bf{H}$ tokamak plasmas}

\author{Ye.O.~Kazakov$^{1}$, T.~F\"ul\"op$^{1}$ and
D.~Van Eester$^{2}$}

\address{
$^1$ Department of Applied Physics, Nuclear Engineering,
Chalmers University of Technology and Euratom-VR Association, G\"oteborg, Sweden \\
$^2$ LPP-ERM/KMS, Association Euratom-`Belgian State', TEC Partner, Brussels, Belgium
}
\ead{kazakov@chalmers.se}

\begin{abstract}
Hydrogen majority plasmas will be used in the initial non-activated
phase of ITER operation. Optimizing ion cyclotron resonance heating
(ICRH) in such scenarios will help in achieving H-mode in
these plasmas.  Past JET experiments with the carbon wall revealed a
significant impact of intrinsic impurities on the ICRH performance in
$\HeH$ plasmas relevant for the full-field initial ITER phase.  
High plasma contamination with carbon impurities
resulted in the appearance of a supplementary mode conversion layer
and significant reduction in the transition concentration of $\He$
minority ions, defined as the concentration at which the change from
minority heating to mode conversion regime occurs. In view of the
installation of the new ITER-like wall at JET, it is important to
evaluate the effect of $\Be$ and $\Wo$ impurities on ICRH scenarios in
$\HeH$ plasmas.  In this paper, an~approximate analytical expression
for the transition concentration of $\He$ minority ions is derived as
a function of plasma and ICRH parameters, and accounting for typical
impurity species at JET. The accompanying 1D wave modeling
supports the analytical results and suggests a~potential
experimental method to reduce $\He$ level
needed to achieve a~specific heating regime by puffing a small amount
of $^{4}{\rm He}$ ions additionally to $\HeH$ plasma.
\end{abstract}

\maketitle

\section{Introduction}
\label{sect:intro}

Ion cyclotron resonance heating (ICRH) has been used successfully for
bulk ion and electron heating in tokamaks, and is foreseen as one of
the additional heating systems to be installed at ITER. At the initial
stage of ITER operation predominantly hydrogen (H) or helium-4
($^4$He) plasma will be used to minimize the activation of the tokamak
components~\cite{mayoral.2011}.  The existing scalings suggest the
H-mode power threshold to be higher by a factor of two for hydrogen
plasmas than for deuterium, and therefore the access to H-mode of
operation is not assured for H plasmas in ITER with the heating powers
that will be available~\cite{gohil.2009.nf}.  Thus, it is particularly
important to optimize the efficiency of radio frequency (RF) heating
for the scenarios relevant for this experimental stage of ITER.

It is well known that for efficient ICRH heating at the fundamental
ion cyclotron (IC) frequency plasmas consisting of at least two
different ion species need to be used. The concentration of one ion
species (majority) is usually much higher than the concentration of
another species (minority ions).  Depending on the relative
concentrations of the species, two heating regimes are usually
identified: minority ion heating (MH) and mode conversion (MC).  MH
requires relatively low minority concentrations, less than some
critical value~\cite{wesson}.
In this regime the majority ions assure
favourable polarization of the fast Alfv\'en wave (FW) launched by the
ICRH antenna at the region of the fundamental cyclotron resonance of
minority ions, which absorb the RF energy and transfer it to bulk
plasma ions and electrons via Coulomb collisions.  Whether indirect
bulk ion or electron heating dominates, depends on the ratio between
the minority tail energy and the critical energy, $E_{\rm
  crit}$~\cite{stix}.  When the averaged energy of the fast minority ions is
above $E_{\rm crit}$ electrons are predominantly heated by collisions
with the fast ions, whereas for the opposite case indirect bulk ion
heating is observed. With the gradual increase in the minority
concentration the MH efficiency reduces and at large enough minority
concentrations plasma heating via MC becomes dominant.  This regime is
characterized by a partial conversion of the FW to the short
wavelength modes, ion Bernstein wave (IBW) and ion cyclotron wave
(ICW), at the MC layer~\cite{jaeger.icw}. The converted wave is
commonly strongly absorbed by electrons within a~narrow spatial region
on much shorter time scale than the characteristic time for indirect
bulk plasma heating via MH.  Thus, as one of the methods to identify
experimentally whether MH or MC heating occurs, is to study electron
temperature response to ICRH power modulation~\cite{mantsinen.2004,
  lerche.2008}. Beyond plasma heating itself, the MC regime has a
number of promising applications in present-day and future fusion
machines, e.g. driving non-inductive current and generating plasma
rotation~\cite{majeski.1996.cd, becoulet.1996, parisot.2007,
  lin.2008.prl, lin.2009.pop, hellsten.2012.ppcf}.

A number of experiments were performed at JET and ASDEX-U aimed at
studying various ICRH heating schemes in H majority plasmas with
helium-3 $(\He)$ minorities that can be used in the non-activated
phase of ITER~\cite{noterdaeme.1999, mayoral.2006, lamalle.2006,
  dve.2012, lerche.2012}.  Both $\HeH$ schemes relevant for the
half-field $(B_0=2.65\,{\rm T})$ and full-field $(B_0=5.3\,{\rm T})$
operation phase of ITER were tested recently. For the frequency range
designed for the ITER ICRH system $(f=40-55\,{\rm MHz})$ two heating
scenarios are feasible for the half-field H phase: fundamental IC
heating of H majority ions at $f \approx 40\,{\rm MHz}$ and second
harmonic heating of $\He$ minority ions at $f \approx 53\,{\rm
  MHz}$~\cite{lerche.2012}. However, both of these scenarios have
relatively low single-pass absorption with dominant fast wave electron
heating.  For the full-field H phase fundamental ion cyclotron heating
of $\He$ ions is the only scheme available for plasma heating, which
commonly has an as good heating efficiency as any of fundamental
minority ICRH scenarios. This scheme has been tested at
JET~\cite{mayoral.2006, lamalle.2006} by adopting the magnetic field
$(B_0 \approx 3.6\,{\rm T})$ and ICRH frequency $(f \approx 37\,{\rm
  MHz})$ to locate $\He$ cyclotron layer in the plasma center as it
will be in ITER.

In JET experiments reported in Refs.~\cite{mayoral.2006, lamalle.2006}
$\HeH$ plasma heating was studied at very low $\He$
concentrations. Mode conversion was found to be reached at $\He$
concentrations $X[\He]=n_{\He}/n_e=2-3\%$, which were substantially
smaller than the values observed in $\HeD$ plasmas ($X[\He] \approx
10-15\%$).  Such a difference is explained by the fact that $\HeH$ is
the so-called `inverted' ICRH scenario.  A key feature of the inverted
scenarios is that the minority ion species have a smaller
charge-to-mass ratio than the majority species, $(Z/A)_{\rm mino} <
(Z/A)_{\rm majo}$.  For these heating scenarios the MC layer is
located between the ICRH antenna on the low-field side (LFS) and the
minority cyclotron resonance (the FW encounters the MC layer first),
while for standard scenarios -- like in $\HeD$ plasmas -- the minority
cyclotron layer is located between the MC layer and the LFS antenna.

Those experiments highlighted an essential effect of impurities in the
inverted $\HeH$ scenario. Three important issues due to plasma
dilution with the carbon impurities were outlined (note that the
experiments \cite{mayoral.2006, lamalle.2006, dve.2012, lerche.2012}
were carried at JET with the inner vessel covered with carbon tiles).
First, the heating region was found to be shifted appreciably away
from where it was expected for pure plasma.  Second, MC heating was
complicated further through the appearance of the supplementary MC
layer associated with carbon (C) impurities.  The efficiency of ICRH
heating in $\HeH$ JET plasmas with multiple MC layers 
was addressed in detail in Ref.~\cite{dve.2012}
for the extended range of $\He$ concentrations,
showing in addition the complexity
of the real time control of the minority level and, thus,
difficulty in controlling the location of the MC layer in such plasmas.
The third direct effect of carbon impurities on ICRH
performance was the reduction of the transition $\He$ minority
concentration, $\Xcrhe$.  Full-wave ICRH simulations have shown that
for the plasma without the carbon the transition from MH to MC heating
should occur at $X_{\rm crit}[\He] \approx 5\%$, while the
experimentally observed levels at JET were lower \cite{mayoral.2006}.

Since August 2011 JET is operating with the new ITER-like wall, using
beryllium (Be) and tungsten (W) as the new plasma facing materials.
Therefore, it is instructive to assess and analyze the impact of
modest amounts of first wall material impurities, which will enter the
plasma due to the plasma-wall interaction, on the performance of ICRH
heating. The aim of the present paper is to find a reasonable estimate
for $X_{\rm crit}[\He]$ in $\HeH$ plasma, which corresponds to the
change of the heating regime from MH to MC, and evaluate the effect of
typical impurities at JET on that.

The paper is organized as follows. Section~\ref{sect:xcrit.num} shows
1D numerical results for the dependence of the transition $\He$
concentration on the plasma and ICRH parameters in pure $\HeH$
plasma. We present two equivalent approaches for the analytical
estimate of $X_{\rm crit}[\He]$ in section~\ref{sect:xcrit.theory}.
In section~\ref{sect:xcrit.imps} one of the approaches is generalized
further to account for the impurities in the plasma.  Also, the
dependence of $X_{\rm crit}[\He]$ on the concentrations of Be and
other impurities is analyzed there.  Based on the results of previous
sections, in section~\ref{sect:proposal} we suggest a potential method
to reduce and/or control $\Xcrhe$ by using additional puffing of
$^{4}{\rm He}$ ions to $\HeH$ plasma. Finally, conclusions are drawn
in section~\ref{sect:concls}.

\section{Numerical results for the transition concentration of
helium-3 ions in pure $\HeH$ plasma}
\label{sect:xcrit.num}

To analyze the wave propagation and damping dynamics in the $\HeH$
plasma, the 1D ICRH full-wave code TOMCAT~\cite{dve.1998} has been
used.  This code solves a 12th order wave equation system, that
guarantees a positive definite and purely resonant absorption for
Maxwellian populations, accounting for the radial
variation of the toroidal magnetic field and FW parallel wavenumber.
It,~however, omits the finite poloidal magnetic field
effects, and thus excludes MC of the FW to the ICW.  All the results
reported assume pure excitation of the FW from the LFS of the tokamak
as the imposed boundary conditions.  TOMCAT gives scattering
coefficients (reflection, transmission, conversion and absorption) for
a single or a double transit of the FW over the plasma depending on a
chosen radial range of integration.  In~contrast to usual full-wave
codes, where all the RF power launched into the plasma is assumed to
be absorbed (i.e. yielding multi-pass absorption), TOMCAT calculates
the single- or double-pass absorption coefficients.  The evaluation of
a single- or a double-pass absorption coefficient allows for
estimating the heating efficiency of the studied ICRH scenario
qualitatively.

Even though TOMCAT can give only qualitative results (since many effects
are not taken into account in a 1D geometry), it is helpful for
understanding the global trends and some of the observed
characteristics of the ICRH performance.  TOMCAT simulations were used
for the analysis of the past $\HeH$ experiments at JET and helped to
identify the effect of C impurities in these
scenarios~\cite{mayoral.2006, dve.2012}.  In a recent
paper~\cite{kazakov.2012.nf}, the normalized absorption coefficients
given by TOMCAT and the evaluated deuterium transition concentration
in D-T plasma were found to be in a reasonable agreement with the
results of more sophisticated modelling with the 2D full-wave code
TORIC. We use therefore TOMCAT modelling to check and supplement the
analytical estimates presented in this paper.  However, it is
important to note that a rigorous treatment of the FW propagation and
MC in tokamaks should be essentially based on the 2D or 3D full-wave
modelling~ \cite{selfo, toric, scenic, aorsa, eve, budny.2012}.

\begin{figure}[htbp]
\centering
\includegraphics[trim=0.0cm 0.0cm 0cm 0.0cm, clip=true, width=7.5cm]{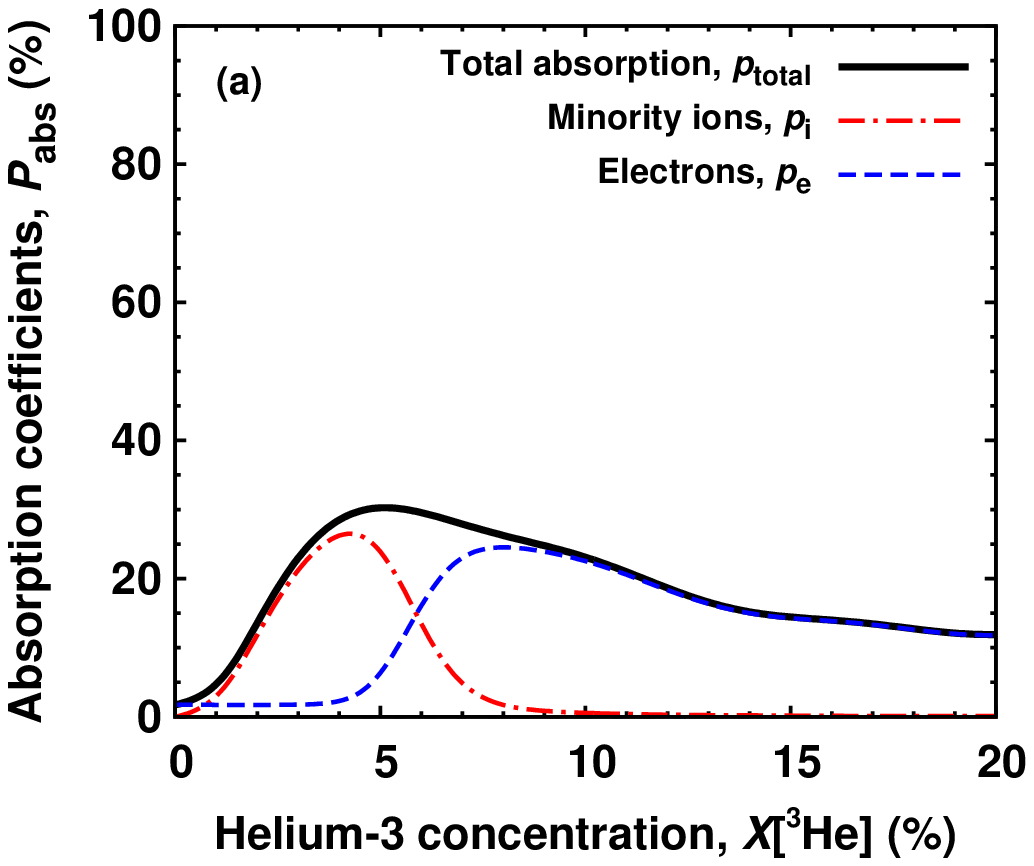}
\includegraphics[trim=0.0cm 0.0cm 0cm 0.0cm, clip=true, width=7.5cm]{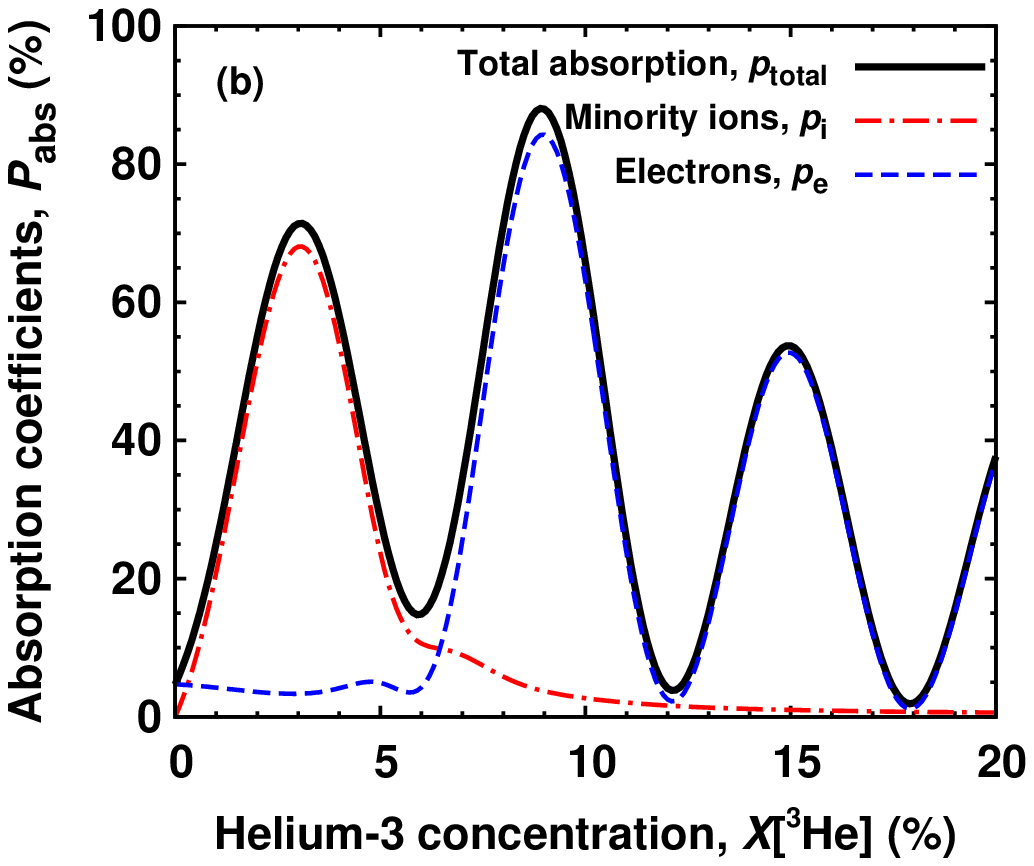}
\caption{Single-pass (a) and double-pass (b) absorption coefficients
vs. $\He$ minority concentration calculated with the TOMCAT code for $\HeH$ plasma: $B_0 = 3.1\,{\rm T}$, $f = 32.2\,{\rm MHz}$, $n_{e0}=3.2 \times 10^{19}\,{\rm m}^{-3}$, $T_0=5.0\,{\rm keV}$, $n_{\rm tor}=27$.}
\label{fig:pabs.vs.x3he}
\end{figure}

Figure~\ref{fig:pabs.vs.x3he} shows the single-pass (left) and
double-pass (right) absorption coefficients in $\HeH$ plasma, computed
with TOMCAT.  We consider plasma and ICRH parameters typical for past
JET $\HeH$ experiments: plasma major and minor radius $R_{0} =
2.96\,{\rm m}$ and $a=0.9\,{\rm m}$, RF frequency $f = 32.2\,{\rm
  MHz}$, central magnetic field $B_0=3.1\,{\rm T}$,
$n_{e}(r)=(n_{e0}-n_{e1})(1 -(r/a)^2) + n_{e1}$, central density
$n_{e0}=3.2 \times 10^{19}\,{\rm m}^{-3}$, edge density
$n_{e1}=0.1\,n_{e0}$, $T(r)=(T_{0}-T_{1})(1 -(r/a)^2)^{1.5} + T_{1}$,
central electron and ion temperature $T_{0}=5.0\,{\rm keV}$, edge
temperature $T_{1}=0.1\,T_{0}$, FW toroidal mode number is taken to be
$n_{\rm tor}=27$ $(k_{\|}= n_{\rm tor}/R)$, typical for dipole phasing
of the A2 ICRH antenna at JET. This choice of $f$ and $B_0$ places
$\He$ cyclotron resonance almost centrally, $R_{\He}=2.9\,{\rm
  m}$. %,$\sim6\,{\rm cm}$ to the HFS of the plasma core.
 The results we present in this section are computed assuming pure
 plasma without impurities.

For the considered parameters single-pass absorption by minority
species reaches its maximum $p_{i}=27.5\%$ at $\Xhe=4.2\%$. Electron
absorption at this minority concentration is only $p_{e}=2.7\%$. Ion
absorption starts to degrade with increasing $\Xhe$, whereas electron
heating via mode conversion, in contrast, increases.  At~$\Xhe=5.9\%$
minority ion heating is balanced by electron heating, and we refer to
that $\Xhe$ as a transition concentration, and denote it with
$X_{\crit}[\He]$. The maximum electron heating $p_{e}=25.6\%$ is
reached at $\Xhe=8.1\%$, and starts to decrease for higher $\Xhe$ in
agreement with the Budden theory for the isolated MC
layer~\cite{budden}.

Figure~\ref{fig:pabs.vs.x3he}(b) shows the absorption power fractions
transferred in a double sweep of the FW in the plasma. The FW is
followed from the incidence side (LFS) and is allowed to reflect once
on the high-field side (HFS) cutoff.  It exhibits oscillations in the
double-pass absorption by electrons and significant increase in the
absorption by $\He$ ions in the MH regime.
% ($p_i=68.1\%$ at $\Xhe=3.1\%$).
These are due to the multiple FW reflections in the plasma and
constructive/destructive interference undergone by the reflected
waves~\cite{majeski, fuchs, kazakov.2010.ppcf, kazakov.2012.nf}.
Electron heating via mode conversion starts to dominate over minority
ion heating at $\Xhe=6.4\%$.  The total phase difference, which
defines the resulting double-pass absorption by ions and electrons,
includes the terms due to the FW reflection on the minority cyclotron
resonance and MC layer, respectively. In general, it is different for
minority and MC heating, and thus the discussed transition
concentration $X_{\crit}[\He]$ differs somewhat from the value
calculated for a single transit of the FW in the plasma.  Since the FW
interference complicates the analysis of the wave propagation and
absorption, and the main emphasis of this paper is to sort out the
effect of impurities on $X_{\crit}[\He]$, the numerical results
presented in the paper rely on treating the propagation of the single
incident wave and ignoring possible additional FW reflection from the
HFS cutoff and supplementary MC layers (single-pass TOMCAT
calculations) unless otherwise stated.

\begin{figure}[htbp]
\centering
\includegraphics[trim=0.0cm 0.0cm 0cm 0.0cm, clip=true, width=7.5cm]{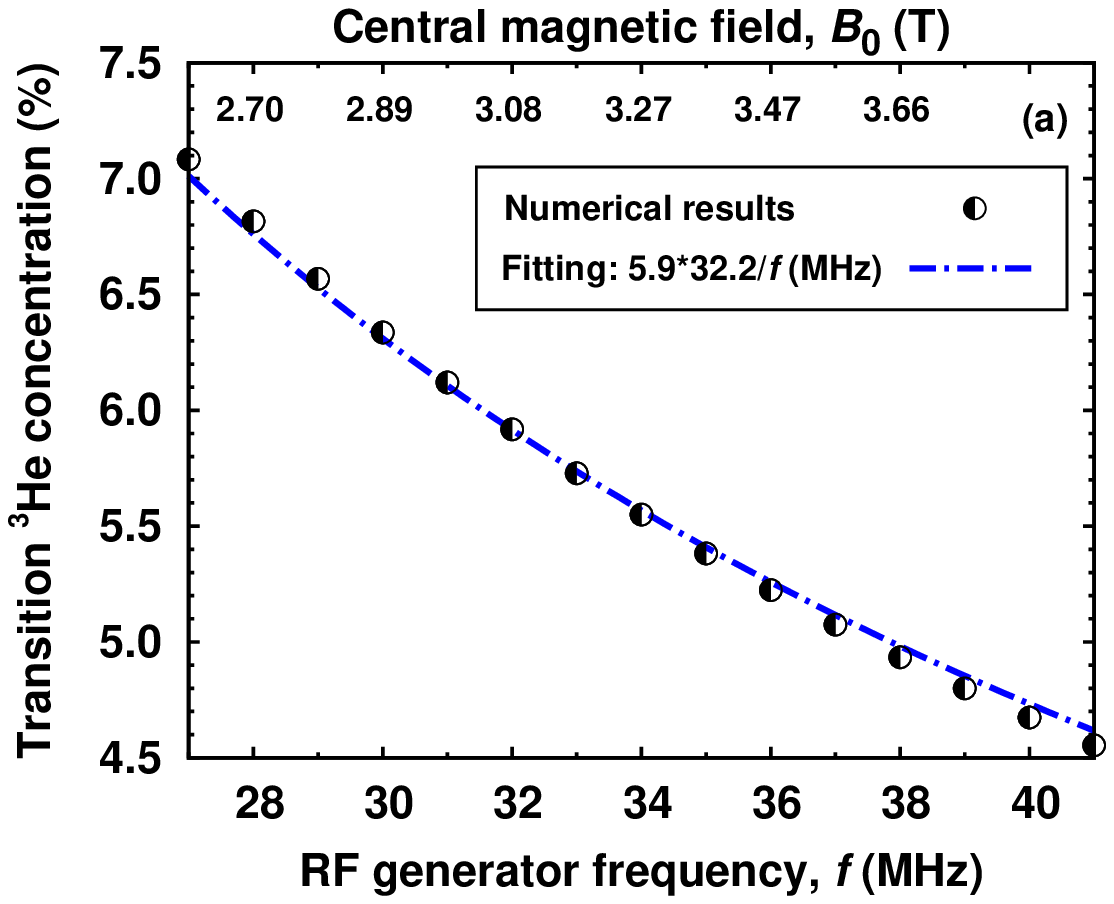}
\includegraphics[trim=0.0cm 0.0cm 0cm 0.0cm, clip=true, width=7.5cm]{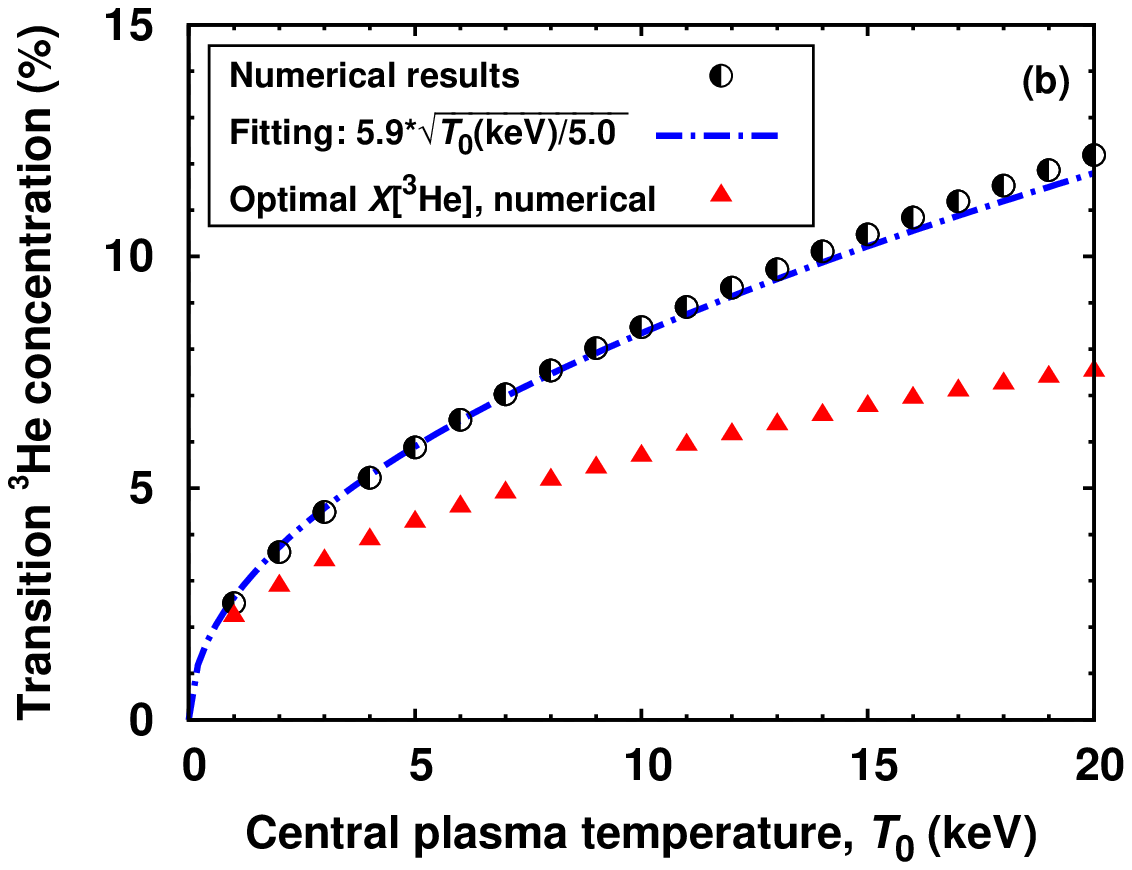}
\includegraphics[trim=0.0cm 0.0cm 0cm 0.0cm, clip=true, width=7.5cm]{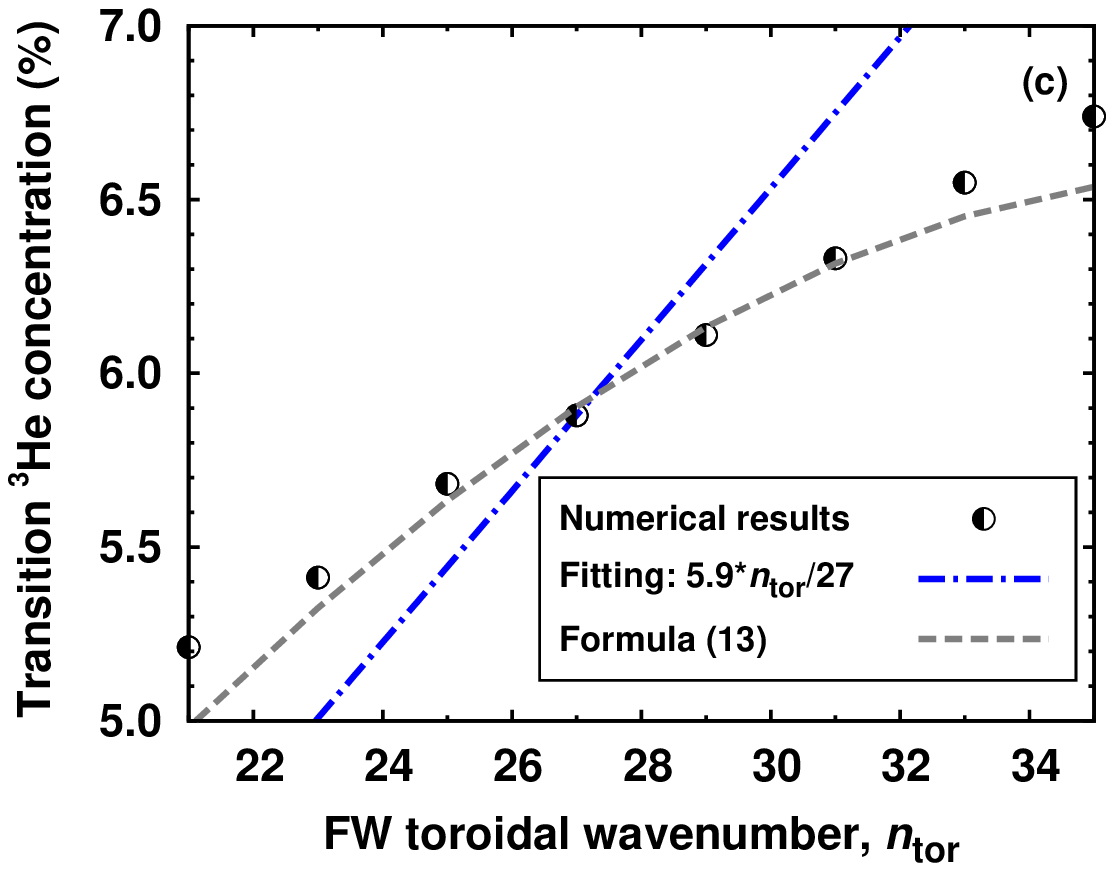}
\includegraphics[trim=0.0cm 0.0cm 0cm 0.0cm, clip=true, width=7.5cm]{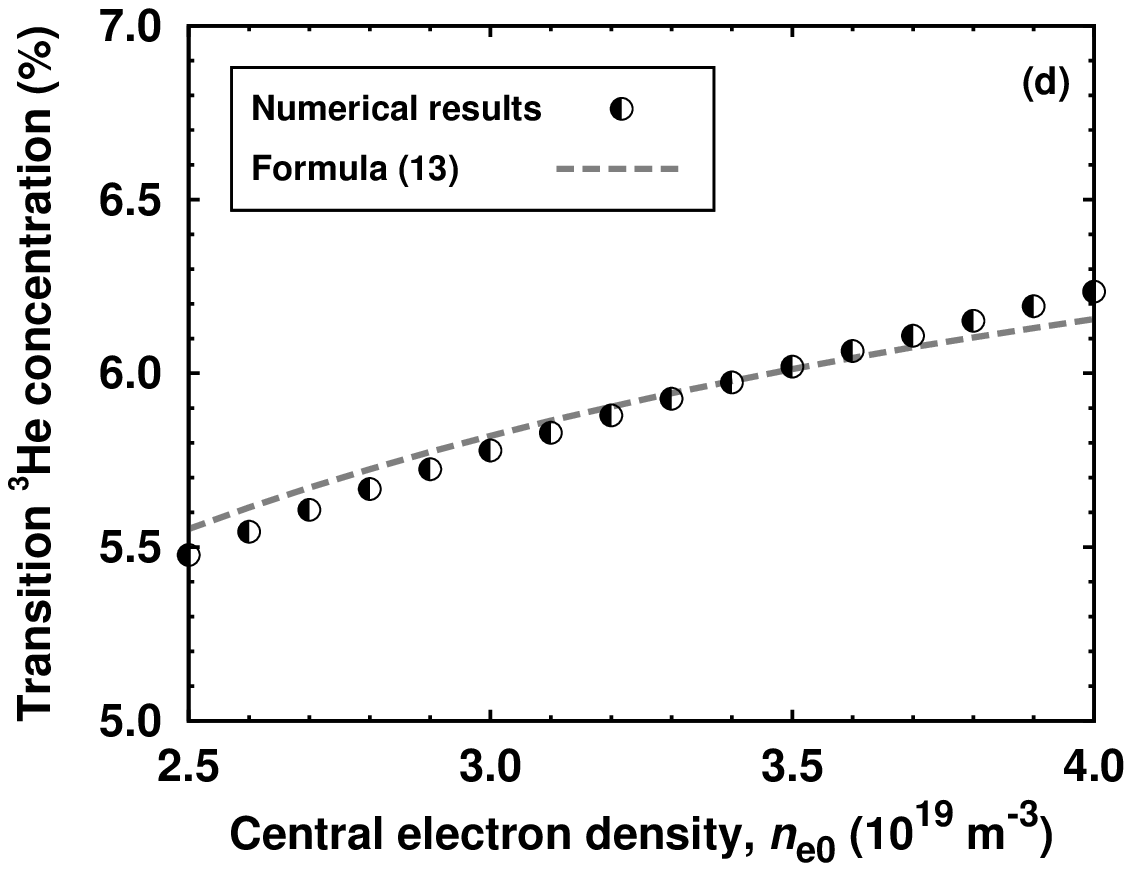}
\caption{Transition concentration of $\He$ ions as a function of
RF frequency ($f/B_0$~fixed) (a), central plasma temperature (b),
FW toroidal wavenumber (c) and central plasma density (d).}
\label{fig:xcrit.vs.param}
\end{figure}

ICRH system on JET covers a frequency range
from 23 to 57~MHz, providing access to a large number of
scenarios for a wide range of magnetic fields~\cite{JET.heating}.
In~$\HeH$ ICRH experiments reported in \cite{mayoral.2006, dve.2012}
different RF frequencies were used for plasma heating, $f \approx
37\,{\rm MHz}$ and $f \approx 32\,{\rm MHz}$, respectively.
Figure~\ref{fig:xcrit.vs.param}(a) presents $X_{\crit}[\He]$ as a
function of RF generator frequency $f$, but for the fixed $f/B_0$
ratio to keep the same location of the minority cyclotron resonance in
the plasma, $R_{\He}$. The transition concentration $X_{\crit}[\He]$
decreases if operating at higher RF frequency and magnetic field:
while for baseline case shown in Fig.~1 ($f/B_0=32.2/3.1$) transition
from MH to MC was reached at $X[\He]=5.9\%$, $\Xcrhe$ decreases to
$5.1\%$ if choosing $f/B_0=37.4/3.6$.  $\Xcrhe$ follows approximately
a $1/f$-dependence, which is clearly seen by comparing the
numerical results with the fitting curve. It suggests that the
transition concentration is connected to the Doppler width of the
minority IC resonance, $\Delta R \propto \sqrt{2} k_{\|} v_{th,
  \rm{mino}} / \omega$, which is also inversely proportional to the RF
frequency.

This interpretation is supported by Fig.~\ref{fig:xcrit.vs.param}(b),
where the dependence of $\Xcrhe$ on the plasma temperature is depicted
(all plasma species are assumed to have equal temperatures). The
transition concentration of $\He$ ions raises with the temperature and
scales as a square root of $T_0$.  In future machines like ITER, where
higher plasma temperatures are expected for the tokamak operation, the
transition to MC in $\HeH$ plasmas is to occur at higher $\He$
concentrations than in JET. Figure~\ref{fig:xcrit.vs.param}(b) also
shows that $X[\He]$, at which single-pass minority ion absorption is
maximized (triangles), increases with $T_0$.  Due to increased $\He$
demand and industrial consumption, the typical market price for $\He$
has raised from \$100 -- \$200 per liter to \$2,000 per liter in
recent years~\cite{he3.price}.  Along with the fact that the plasma
volume in ITER is almost 10 times larger than in
JET, %$(V_{\rmpl}=840\,{\rm m}^{3})$,
this increases significantly the operational costs for using $\He$ in
future fusion devices.  The present paper discusses the
  possibility to retune ICRH scenarios involving helium-3 ions to
  minimize $\He$ concentrations needed for MH and MC heating.

As follows from Fig.~\ref{fig:xcrit.vs.param}(c), the transition
concentration increases with the FW toroidal wavenumber, which is
consistent with the increase in the Doppler broadening of the
cyclotron resonance for higher $k_{\|}$. However, the calculated
values of $\Xcrhe$ are less sensitive to the value of $n_{\rm tor}$
than the estimates yielded from the linear fitting. A~moderate
increase in $\Xcrhe$ is also observed if increasing the central plasma
density, as Fig.~\ref{fig:xcrit.vs.param}(d) illustrates. The reason
why $\Xcrhe$ raises with $n_{e0}$ will be outlined in the next
section, where we present simplified analytical models that help us to
grasp the main physical parameters affecting $\Xcrhe$.

% http://spectrum.ieee.org/biomedical/diagnostics/physics-projects-deflate-for-lack-of-helium3

\section{Analytical estimates for the transition concentration of
helium-3 ions in pure $\HeH$ plasma}
\label{sect:xcrit.theory}

ICRH heating of fusion plasmas relies on the transport of the energy
by the fast waves from the edge of the plasma to the core. The FW
propagating at the equatorial plane of a tokamak
is fairly well described by the dispersion
relation
\beq
n_{\perp, {\rm FW}}^2 = \frac{(\epsL-\npar^2)(\epsR-n_{\|}^2)}{\epsS-n_{\|}^2},
\label{eq:fw.disp}
\eeq
where $\npar = c \kpar / (2\pi f)$ is the refractive index parallel to
the equilibrium magnetic field, and $f$ is the RF generator (antenna)
frequency.  Omitting the poloidal magnetic field, the FW parallel
wavenumber varies radially as $\kpar = n_{\rm tor}/R$; the dominant
$n_{\rm tor}$ in the launched FW spectrum is determined by the antenna
geometry and chosen antenna phasing. In Eq.~(\ref{eq:fw.disp}),
$\epsS$, $\epsL$ and $\epsR$ are the plasma dielectric tensor
components in the notation of Stix~\cite{stix}, which in a cold-plasma
limit and for the ICRH frequency range are given by \beq \eqalign{
  \epsS \approx 1 + \frac{\ompe^2}{\omce^2} -\sum_{i} \frac{\ompi^2}{\om^2 - \omci^2}, \\
  \epsL \approx 1 + \frac{\ompe^2}{\omce^2} +\sum_{i} \frac{\ompi^2}{\omci(\omci - \om)}, \\
  \epsR \approx 1 + \frac{\ompe^2}{\omce^2} +\sum_{i}
  \frac{\ompi^2}{\omci(\omci + \om)},
\label{eq:tensor.comps}
}
\eeq
where the summation is to be taken over all ion species constituting the plasma,
and $\omega_{ps}$ and $\omega_{cs}$ are the species' plasma and cyclotron
frequencies, respectively.

The FW resonance condition $\epsS = \npar^2$ defines the location of
the MC layer, often called also as the ion-ion hybrid (IIH)
resonance. Hot plasma theory resolves this resonance and bends it into
a confluence.  At this layer the FW is converted partially to a short
wavelength mode, the IBW and the ICW, depending on the relation
between the plasma temperature and poloidal magnetic field. The MC
layer is accompanied closely to the LFS by the left-hand polarized
cutoff (L-cutoff) defined by the condition $\epsL = \npar^2$. In
Appendix~A of the paper we present the derivation of the formulas for
resonant frequencies, $\ws$ and $\wl$, satisfying the conditions \beq
\eqalign{
%\left. \epsS \right|_{\omega = \ws} = \npar^2, \\
\epsS = \npar^2, \\
\epsL = \npar^2,
\label{eq:res.conds}
} \eeq which we refer to further as the IIH and L-cutoff frequencies,
respectively.

Since the toroidal magnetic field follows $1/R$ dependence in
tokamaks, the location of the layer, at which the antenna frequency
$\omega$ matches a certain resonant frequency $\omega_{i}$ $(\omega =
2\pi f = \omega_{i})$, can be calculated as follows \beq \left. R
\right|_{\omega = \omega_{i}} = R_{0} \,
\frac{\omega_{i}}{\omega_{c{\rm H}}} \, \frac{15.25 \,B_{0}(\rm T)}{f
  \rm(MHz)},
\label{eq:Ri.vs.param}
\eeq where $\omega_{c{\rm H}}$ is the cyclotron frequency for hydrogen
ions. As an example, for the baseline conditions chosen ($f=32.2\,{\rm
  MHz}$, $B_0=3.1\,{\rm T}$) the hydrogen IC resonance is located at
$R_{\rm H}=4.34\,{\rm m}$, $\He$ resonance -- at $R_{\He}=2.90\,{\rm
  m}$, $\rm{Be}$ resonance -- at $R_{\rm Be}=1.93\,{\rm m}$.  Vice
versa, using Eq.~(\ref{eq:Ri.vs.param}) we can connect the radial
coordinate in the plasma, $R_i$ to the corresponding resonant
frequency, $\omega_{i}$.

%The numerical coefficient in the numerator of Eq.~(4) is
%equal to $e/(200\pi c m_{\rm p} )$

For the discussion of the transition minority concentration and in
order to understand how two ICRF heating regimes arise in two-ion
component plasmas, we need to account for the kinetic response of the
resonant minority ions in the $\epsS$ tensor component in the
denominator of Eq.~(\ref{eq:fw.disp})~\cite{wesson}.  To keep the
algebra simpler, we consider the leading order terms in finite Larmor
radius expansion of $\epsS$. Then, it can be written as follows \beq
\epsS = 1 + \frac{\omega_{pe}^2}{\omega_{ce}^2} + \sum_{i=1,2}
\frac{\ompi^2}{2\om^2} \ffraci \lpb \zzetai{1} + \zzetai{-1} \rpb,
\label{eq:epsS.hot}
\eeq where $Z(\xi)$ is the plasma dispersion function,
$v_{ti}=(T_{i}/m_{i})^{1/2}$ is the thermal velocity, and $\xi_{\pm
  1i}=(\om \pm \omci )/(\sqrt{2} \kpar v_{ti} )$.  Following
Ref.~\cite{wesson}, throughout the paper indices `1' and `2' refer to
majority (hydrogen) and minority (helium-3) ions, respectively.
Majority species are non-resonant, so one could use asymptotic
expansion of the plasma dispersion function for large arguments,
$Z(\xi) \simeq - 1/\xi$. For minority ions the term proportional to
$Z(\xi_{12})$ is non-resonant, while the term proportional to
$Z(\xi_{-12})$ represents resonant minority response. In such a way,
we could write $\epsS$ tensor element as \beq \epsS \approx -
\frac{\omp{1}^2}{\om^2 - \omc{1}^2} - \frac{\omp{2}^2}{2 \om (\om +
  \omc{2})} + \frac{\omp{2}^2}{2\om^2} \ffrac{2} Z(\xi_{-12}).
\label{eq:epsS.hot2}
\eeq

\begin{figure}[htbp]
\centering
\includegraphics[trim=0.0cm 0.0cm 0cm 0.0cm, clip=true, width=13cm]{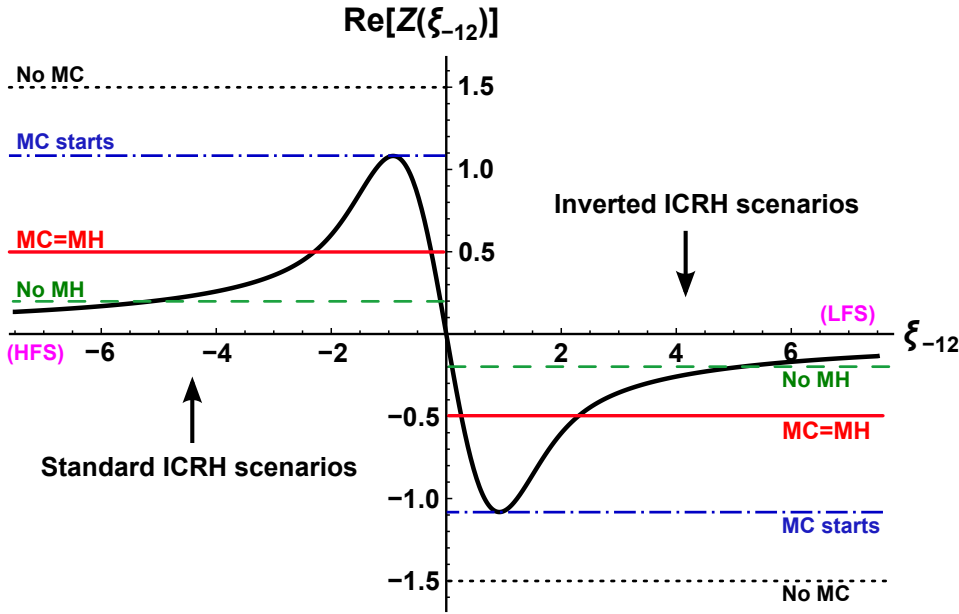}
\caption{MC occurs when there is an intersection of the horizontal
line, which represents the right-hand side of Eq.~(\ref{eq:Zfunc.eq.rhs}), with
the real part of the plasma dispersion function, $Z(\xi_{-12})$. For the standard/inverted ICRH scenarios the resonance condition $\epsS=\npar^2$ is fulfilled to the HFS/LFS
of the minority IC layer, respectively.}
\label{fig:Zfunc}
\end{figure}

Accounting for $\omega \approx \omega_{c2}$ and that the
minority concentration is typically much lower than
that of majority level,
the resonance condition $\epsS = \npar^2$
can be re-written
\beq
Z(\xi_{-12}) = \frac{2\omp1^2}{\omp2^2} \frac{\sqrt{2}k_{\|}v_{t2}}{\omega}
\lpb \frac{\mu^2}{1-\mu^2} + \frac{\kpar^2 v_{A1}^2}{\omega^2} \rpb,
\label{eq:Zfunc.eq.rhs}
\eeq
where we have introduced $\mu = Z_1 A_2/Z_2 A_1$, and $v_{A1}$
is the Alfv\'en speed corresponding to the majority ions.
The real part of $Z(\xi)$ function is shown in Fig.~\ref{fig:Zfunc}; note that the argument
of the plasma dispersion function is proportional to the distance
from the minority cyclotron resonance layer $(R_2=R_0 + x_2)$
\beq
\xi_{-12}(x) = \frac{\omega}{\sqrt{2}k_{\|}v_{t2}}\frac{x - x_2}{R_2}.
\label{eq:xim12}
\eeq
The right-hand side of Eq.~(\ref{eq:Zfunc.eq.rhs}) depends on the
minority concentration $X_2=n_{2}/n_{e}$ via $\omp2^2$ in the
denominator, but is independent of $\xi_{-12}$ and thus is represented
by a~horizontal line in Fig.~\ref{fig:Zfunc}. Note that the second
term in Eq.~(\ref{eq:Zfunc.eq.rhs}) is small compared to the first for
the typical experimental conditions.  The resonance and MC occur
provided there is an intersection of the horizontal line with the
curve for $Z(\xi)$ function.  For the standard ICRH scenarios $\mu <
1$, while for the inverted -- $\mu >1$; therefore, the resonance
condition for MC can be fulfilled only on the HFS and LFS of the
minority IC resonance, respectively.

In Fig.~\ref{fig:Zfunc} we have depicted 4 horizontal lines
corresponding to different minority concentrations. When $X_2$ is very
small, there is no intersection of the horizontal line (dotted) with
${\rm Re}[Z(\xi)]$ and thus the resonance condition for MC does not
occur.  The FW is absorbed by minority ions due to the imaginary part
of $Z(\xi_{-12})$ and gives rise to the MH regime. When, on the
contrary, $X_2$ is very high, there is always an intersection at
$|\xi_{-12}| > 5$, for which the imaginary part of $Z(\xi)$ is
negligible (dashed line).  This is a~MC regime, when the absorption is
defined by the local resonance and there is almost no minority
damping.  When the minority concentration is such that the horizontal
curve is tangential to the real part of $Z(\xi)$, then MC starts to
occur (dash-dotted line). Since the maximum absolute value of ${\rm
  Re}[Z(\xi)]$ is approximately unity, then we can derive the
corresponding minority concentration to be~\cite{wesson} \beq
X_{2}^{\rm{(Wesson)}} = \frac{\sqrt{2}\kpar v_{t2}}{\omega}
\frac{2A_1}{A_2 Z_1} \lpb \frac{\mu^2}{|1-\mu^2|} \pm \frac{\kpar^2
  v_{A1}^2}{\omega^2} \rpb.
\label{eq:x2.wesson}
\eeq The difference between Eq.~(\ref{eq:x2.wesson}) and the result
presented in Ref.~\cite{wesson} is `$\pm$' sign for the second term: plus/minus
sign should be taken for the standard/inverted ICRH scenario,
respectively.  The physical interpretation for this fact is given in
the Appendix~B of the paper.  Since
%$v_{A1}^2/c^2 = \omega_{c1}^2/\omega_{p1}^2 \propto 1/n_{e}$,
$v_{A1}^2 \propto 1/n_{e}$, for the inverted $\HeH$ scenario the
transition minority concentration slightly increases with the plasma
density as it was shown in Fig.~\ref{fig:xcrit.vs.param}(d).  Note
that criterion expressed by Eq.~(\ref{eq:x2.wesson}) was first
indicated by Takahashi~\cite{takahashi}.
In~Ref.~\cite{lashmore-davies}, Lashmore-Davies \etal generalized this
expression for the cases of degenerate resonances in $\HDe$ plasmas
(the majority second harmonics coincides with the minority
fundamental) and a single-species second harmonic heating.

\begin{figure}[htbp]
\centering
\includegraphics[trim=0.0cm 0.0cm 0cm 0.0cm, clip=true, width=11.5cm]{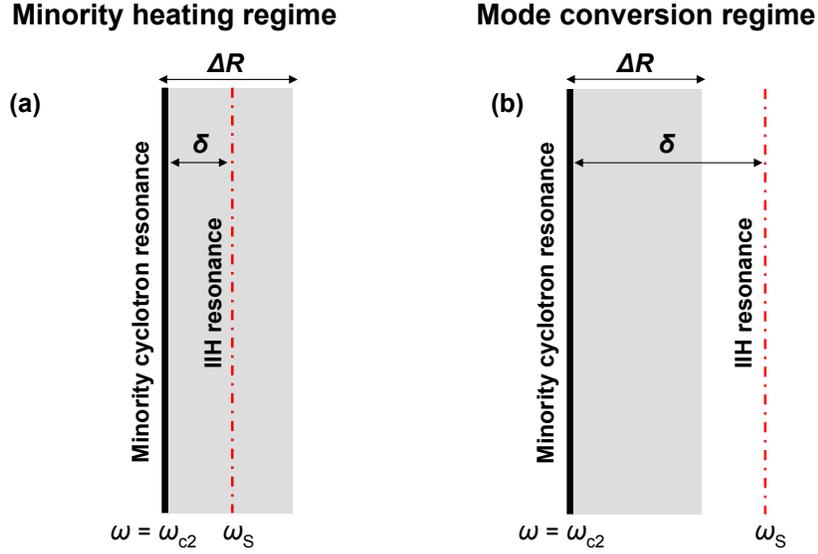}
\caption{Transition from MH (a) to MC (b) heating regime
occurs when the IIH resonance passes through
the Doppler broadened minority cyclotron layer (the scheme
corresponds to the inverted ICRH scenario, for which
the IIH resonance is to the LFS of the minority IC layer).}
\label{fig:mh.to.mc.transition.scheme}
\end{figure}
We can obtain the same result as that given by
Eq.~(\ref{eq:x2.wesson}) using the different approach. Apart from
giving additional physical insight, this approach will make it easier
to include the effect of multiple impurities in the model.
Qualitatively the transition from the minority heating to mode
conversion heating can be explained as follows~(see
Fig.~\ref{fig:mh.to.mc.transition.scheme}). Minority cyclotron
resonance has a finite Doppler width, $\Delta R = {p}_{0} \sqrt{2}
\kpar v_{t2} R/\omega$, where the numerical coefficient $p_0$ is of
the order of unity.  Let denote $\delta$ as a distance between the
cold-plasma IIH resonance and the minority IC layer.  For small
minority concentrations the IIH layer is located within the Doppler
broadened IC resonance (shaded area in
Fig.~\ref{fig:mh.to.mc.transition.scheme}), and minority heating
dominates (Fig.~\ref{fig:mh.to.mc.transition.scheme}(a)). For large
minority concentrations the IIH resonance is located out of the
region, where the cyclotron damping by minority ions is important,
such that $\delta > \Delta R$, and electron heating via mode
conversion will become the main absorption
mechanism~(Fig.~\ref{fig:mh.to.mc.transition.scheme}(b)).  As noted in
Ref.~\cite{lin.2009.pop}, the transition from MH to MC is reached when
$\delta = \Delta R$, i.e.  when the mode conversion layer passes
through the broadened IC resonance.

For the tangential case discussed in Ref.~\cite{wesson}, the
intersection of the curves occurs at $|\xi_{-12}| \approx 1$.  At this
minority concentration, MC only starts to occur and is much weaker
than minority ion damping.  MC is expected to have the same absorption
strength as the ion absorption for higher $X_2$, when $\left| {\rm
  Re}[Z(\xi_{-12})] \right| \simeq 0.5$ (solid line in
Fig.~\ref{fig:Zfunc}).  The argument of the plasma dispersion
function, which is formally the constant $p_0$
(cf.~Eq.~(\ref{eq:xim12})), for this case equals to $\xi_{-12} \simeq
\pm 2.3$.
%The arguments for the Doppler width presented above leave undefined the exact value
%for the constant $k_0$. One can note that for the discussed tangential case, which
%marks the beginning of mode conversion, the intersection
%of the ${\rm Re}\,Z(\xi_{-12})$ curve with the dash-dotted horizontal
%line occurs at $|\xi_{-12}| \approx 1$.
According to Eq.~(\ref{eq:Ri.vs.param}), the radial coordinate $R_2 \pm \Delta R$ corresponds to the
normalized frequency ($\omn = \omega/\omega_{c{\rm H}}$)
\beq
\omn_{\Delta} = \zn_{2} \lpb 1 \pm p_0\,\sqrt{2}k_{\|}v_{t2}/\omega \rpb,
\label{eq:wn.doppler}
\eeq
where $\zn_{i}=Z_{i}/A_{i}$ denotes
the ratio of the charge number to the atomic mass for ion species.

At the same time the normalized IIH frequency (satisfying
the condition $\epsS = \npar^2$)
as a function of
the minority concentration and $\kpar$ is given by (see Appendix~A)
\beq
\omn_{\rm S} \approx \zn_2 + \frac{\zn_1^2 -\zn_2^2}{2\zn_1} f_2 + \frac{(\zn_1^2 - \zn_2^2)^2}{2\zn_1^2}\,\alpha f_2,
\label{eq:wns.pure}
\eeq where we introduced the convenient notation $f_i = Z_i X_i$ for
the fraction of electrons replaced by the ion species `$i$'.  The last
term in Eq.~(\ref{eq:wns.pure}) describes the shift of the IIH
resonance due to finite $n_{\rm tor}$, $\alpha = (\omega_{c{\rm
    H}}^2/\omega_{p{\rm H}}^2) \npar^2$, $\omega_{p{\rm H}}^2=4\pi n_e
e^2/m_{\rm H}$.  The condition $\delta = \Delta R$ (corresponding to
$\omn_{\Delta} = \omn_{\rm S}$) will be fulfilled if \beq X_{2,
  \crit} = p_0 \frac{\sqrt{2}\kpar v_{t2}}{\omega}\frac{2}{A_2} \lpb
\frac{\zn_1}{|\zn_1^2 - \zn_2^2|} \pm \alpha \rpb,
\label{eq:x2.kazakov.general}
\eeq which is the same as $X_{2}^{\rm{(Wesson)}}$ given in
Eq.~(\ref{eq:x2.wesson}) (except the multiplier $p_0$), but written
with the new notation ($\alpha \zn_1 = \kpar^2 v_{A1}^2/\omega^2$).
For the tangential case, which corresponds to the beginning of MC,
$p_0 \approx 1$ and the results given by Eqs.~(\ref{eq:x2.wesson}) and
(\ref{eq:x2.kazakov.general}) are identical.  On the other hand, to
estimate the transition minority concentration $p_0 \approx 2.3$
should be used, yielding for $\HeH$ plasma ($\zn_1=1$, $\zn_2=2/3$)
\beq X_{\crit}[\He] \approx 2.3\, \frac{\sqrt{2}\kpar v_{t2}}{\omega}
\lpa 1.2 - 2\alpha/3 \rpa.
\label{eq:x2.in.3heh}
\eeq An additional argument for choosing the horizontal intersection
level with ${\rm Re}[Z(\xi)]$ curve in Fig.~\ref{fig:Zfunc} to be
lower than unity is due to the good correspondence between the
numerical results and the analytical formula~(\ref{eq:x2.in.3heh})
which is presented in Figs.~\ref{fig:xcrit.vs.param}(c) and~(d).

In this section we considered pure two-ion majority/minority plasma
excluding the presence of impurities.  The critical minority
concentration in diluted plasmas will be upshifted or downshifted
$\tilde{X}_{2, \crit} = M_{\imp} X_{2, \crit}$, if accounting for
impurities that provide the correction factor $M_{\imp}$.  Thus, the
choice of $p_0$ constant affects the absolute value for the transition
concentration, but it is actually not important to describe the
relative influence of impurities on that, which is a subject of the
next chapter.

\section{Transition concentration of
helium-3 ions in $\HeH$ plasma accounting for impurities}
\label{sect:xcrit.imps}

The reason why impurities affect the transition concentration of
minority ions can easily be understood by noting that the location of
the IIH resonance depends on the level of impurity contamination. In
pure plasma it is described by Eq.~(\ref{eq:wns.pure}), but as
impurities are accounted for additional terms arise.  In
Ref.~\cite{kazakov.2010.ppcf}, three-ion component plasmas were
considered, and the correction term describing the effect of a single
impurity species was obtained. In Appendix~A of this paper, we
generalize those formulas to take into account any number of impurity
species.  It is shown, that to the lowest order the contribution to
$\omn_{\rm S}$ caused by impurities is described by the expression
\beq \delta\omn_{\rm S} \approx \sum_{\imp} \frac{(\zn_1 +
  \zn_2)(\zn_2^2 + \zn_1 \zn_{\imp})}{2\zn_1^2 (\zn_2 + \zn_{\imp})}
\frac{(\zn_1 - \zn_2)(\zn_1 - \zn_{\imp})}{(\zn_2 - \zn_{\imp})}
f_{2}f_{\imp}.
\label{eq:wns.imps}
\eeq Note that in the corresponding formula, derived for a single
impurity in Ref.~\cite{kazakov.2010.ppcf}, due to a misprint a
$1/(2\zn_2)$ factor is missing in a term describing the impurity
response.  If we denote $X_{2, \crit}$ to be the transition minority
concentration in a pure plasma, and $\tilde{X}_{2, \crit}$ as that in
plasmas with impurities, then the following approximate formula can be obtained
\beq \eqalign{ \tilde{\omega}_{\rm S} \approx \zn_2 + \frac{\zn_1^2
    -\zn_2^2}{2\zn_1} f_2 + \frac{(\zn_1^2 - \zn_2^2)^2}{2\zn_1^2}\,
  \alpha f_2 \approx \zn_2 + \frac{\zn_1^2 -\zn_2^2}{2\zn_1}
  \tilde{f}_2\,+ \\ +\,\frac{(\zn_1^2 - \zn_2^2)^2}{2\zn_1^2}\, \alpha
  \tilde{f}_2 + \sum_{\imp} \frac{(\zn_1^2 - \zn_2^2)(\zn_2^2 + \zn_1
    \zn_{\imp})(\zn_1 - \zn_{\imp})}{2\zn_1^2(\zn_2^2 - \zn_{\imp}^2)}
  \tilde{f}_2 f_{\imp}.
\label{eq:wnd=wns.imps}
}
\eeq
From Eq.~(\ref{eq:wnd=wns.imps}) it is easy to show that
\beq
\tilde{f}_2/f_{2} \approx 1 -\sum_{\imp} \frac{(\zn_1 - \zn_{\imp})(\zn_2^2 + \zn_1 \zn_{\imp})}{(1+\alpha(\zn_1^2 - \zn_2^2)/\zn_1)\zn_1(\zn_2^2 - \zn_{\imp}^2)} f_{\imp}.
\label{eq:f2.imps1}
\eeq In the denominator of Eq.~(\ref{eq:f2.imps1}), $\alpha(\zn_1^2 -
\zn_2^2)/\zn_1$ can be neglected since
accounting for that gives the contribution $\propto \alpha f_{\imp}$, and a number of
quadratic terms have already been omitted when deriving
Eq.~(\ref{eq:wns.imps}).  These quadratic terms are summarized in the
Appendix~A of the paper. Thus, the transition minority
concentrations in pure and plasmas contaminated with impurities are
related as follows \beq \frac{\tilde{X}_{2, \crit}}{X_{2, \crit}}
\approx 1 - \sum_{\imp} \frac{(\zn_1 - \zn_{\imp})(\zn_2^2 + \zn_1
  \zn_{\imp})}{\zn_1(\zn_2^2 - \zn_{\imp}^2)}\,f_{\imp}.
\label{eq:x2.imps}
\eeq
Since in $\HeH$ plasmas $\zn_1 > \zn_2 > \zn_{\imp}$, the impurity
contamination leads to the reduction of the transition concentration
of helium-3 ions ($M_{\rm imp} < 1$):
\beq
\eqalign{
M_{\imp} = \frac{\tilde{X}_{\crit}[\He]}{X_{\crit}[\He]}\, \approx 1 - 8 X[{\Be}] - 14.6 X[\C] - 33.6 X[\W{28}] - \\
-62.7 X[\W{46}]
- 51.4 X[{\rm Ni}^{26+}] -...
\label{eq:Mimp.imps}
} \eeq Estimates for the concentrations of impurities typical at JET
are summarized in Table~\ref{tab:ximps}. The data has been taken
from Refs.\cite{coenen, czarnecka, dve.eps.2012}. Similarly to the effect
of C ions for the old JET wall, the presence of $\Be$ is expected to
be the main contributing impurity to the reduction factor $M_{\imp}$ for the new ITER-like
wall.

\begin{table}
\caption{\label{tab:ximps} Rough estimates for the impurity concentrations
at JET equipped with the new ITER-like wall.}
\begin{indented}
\item[]\begin{tabular}{c c c c c c}
\br
 & Be$^{4+}$ & C$^{6+}$ & W$^{46+}$ & Ni$^{26+}$ \\
\br
$X_{\imp} $ & $2.5 \times 10^{-2}$ &  $1.5 \times 10^{-3}$ & $1.0 \times 10^{-4}$ & $1.0 \times 10^{-4}$ \\
\ms
\hline
\ms
$f_{\imp}$ & 0.10 & 0.009 & 0.005 & 0.003\\
\ms
\hline
\ms
$\Delta Z_{\rm eff}$ & 0.30 & 0.05 & 0.21 & 0.07 \\
\br
\end{tabular}
\end{indented}
\end{table}

\begin{figure}[htbp]
\centering
\includegraphics[trim=0.0cm 0.0cm 0cm 0.0cm, clip=true, width=7.5cm]{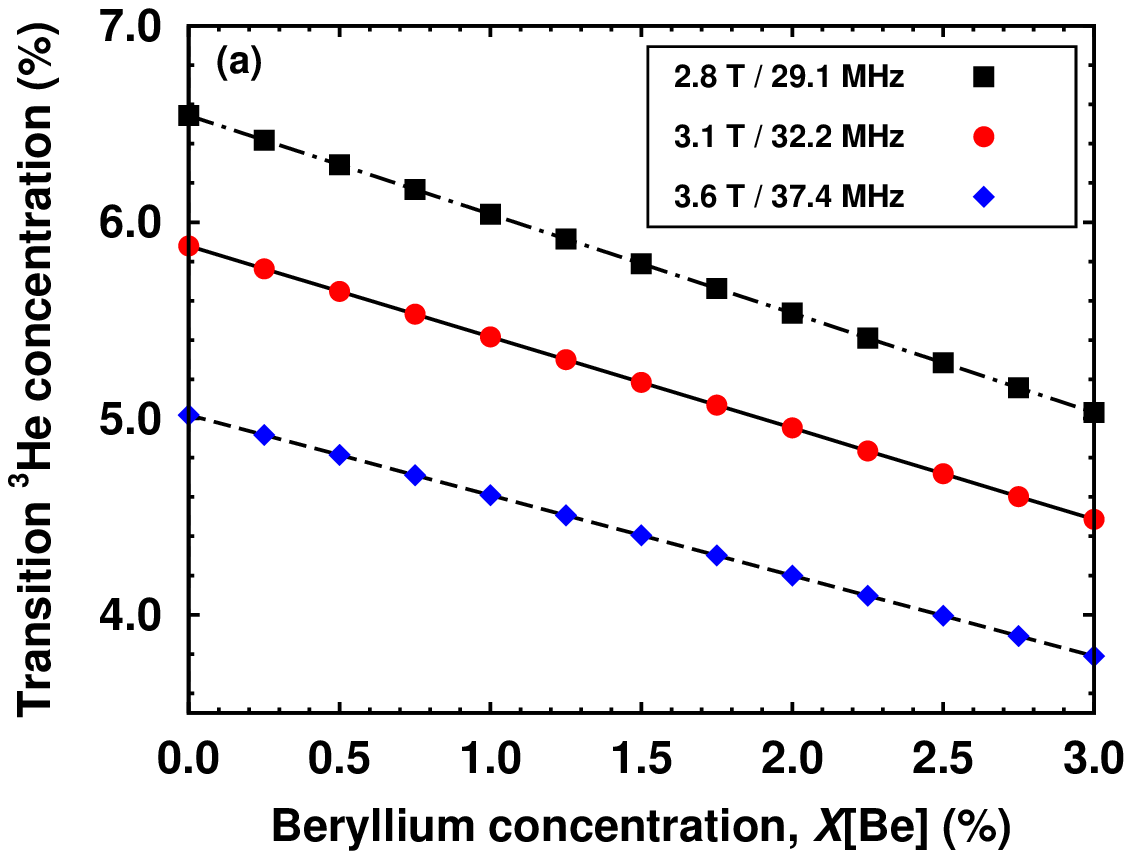}
\includegraphics[trim=0.0cm 0.0cm 0cm 0.0cm, clip=true, width=7.5cm]{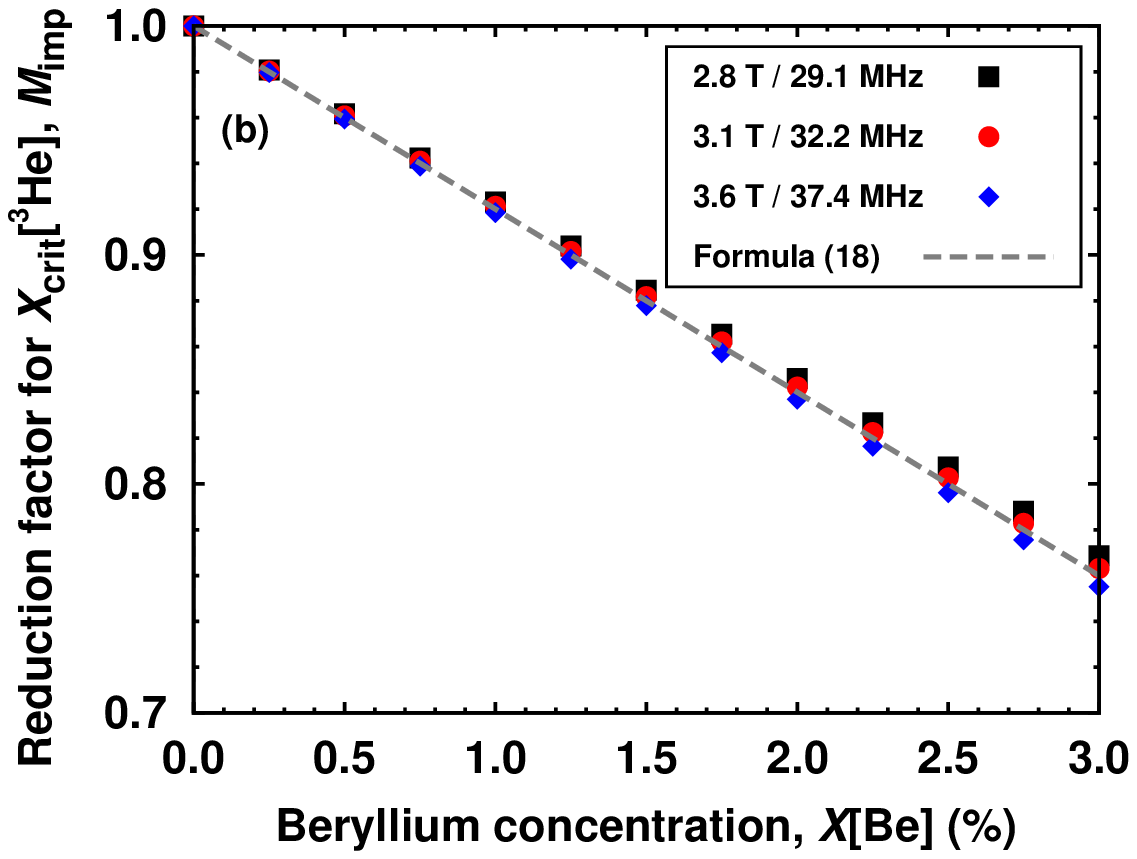}
\caption{(a) Transition concentration of helium-3 ions in $\HeH$ plasma
as a~function of $X[\Be]$ for different $f$ and $B_0$. (b)
Relative change of $\Xcrhe$ is independent of $f$ and decreases
almost linearly with beryllium concentration.
\label{fig:xcrit.vs.be}
}
\end{figure}

Although a number of simplified assumptions have been made to derive
Eq.~(\ref{eq:Mimp.imps}), it gives a remarkably good correspondence
with numerical values.  Figure~\ref{fig:xcrit.vs.be}(a) shows the
transition concentration of $\He$ ions as a function of Be
concentration for different $f$ and $B_0$ ($f/B_0$ fixed), calculated numerically with the TOMCAT solver.  As
discussed, $\Xcrhe$ is smaller if operating at higher antenna
frequencies. Regardless which RF~frequency $f$ is chosen, the transition
concentration decreases with the concentration of $\Be$ impurities.
As shown in Fig.~\ref{fig:xcrit.vs.be}(b), the relative change of
$\Xcrhe$ due to the presence of $\Be$ does not depend on $f$ and is in
a very good correspondence with Eq.~(\ref{eq:Mimp.imps}), $M_{\imp} =
1 - 8 X[\Be]$.

\begin{figure}[htbp]
\centering
\includegraphics[trim=0.0cm 0.0cm 0cm 0.0cm, clip=true, width=7.5cm]{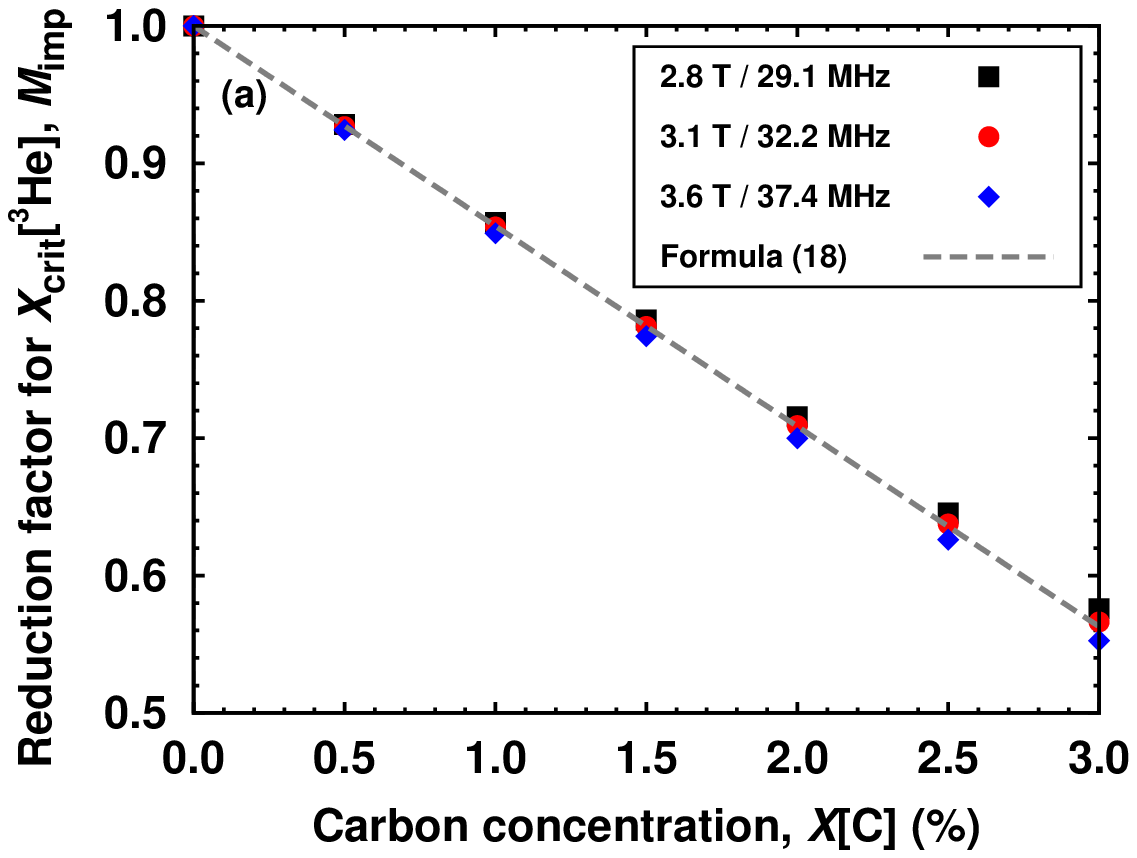}
\includegraphics[trim=0.0cm 0.0cm 0cm 0.0cm, clip=true, width=7.5cm]{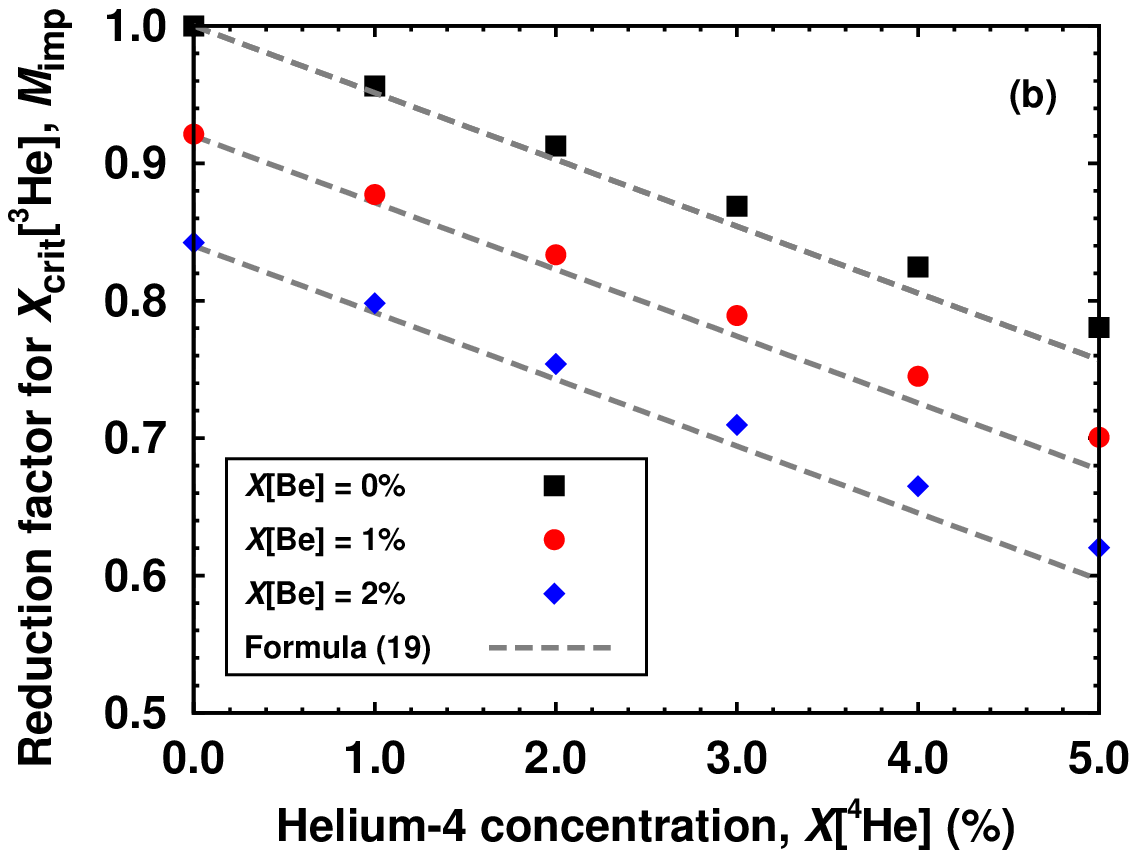}
\caption{Transition concentration of helium-3 ions in $\HeH$ plasma as
  a~function of carbon (a) and helium-4 (b) concentrations. Figure (b)
  is for $f=32.2\,{\rm MHz}$ and $B_{0}=3.1\,{\rm T}$.
\label{fig:xcrit.vs.c.4he}
}
\end{figure}

The numerical factors, which appear in Eq.~(\ref{eq:Mimp.imps}), have
been tested for other impurities, and Fig.~\ref{fig:xcrit.vs.c.4he}(a)
shows an excellent agreement for C ions (relevant for the old JET
wall). The values in Fig.~\ref{fig:xcrit.vs.c.4he} are calculated
numerically with the TOMCAT solver. A rough estimate for $\Be$
concentration at JET, $X[\Be]=2.5\%$ ($\Delta Z_{\rm eff}=0.3$),
should result in the reduction of the transition concentration of
helium-3 ions by about 20\%; it is smaller than the effect of carbon
impurities reported previously ($X[\C]=2.5\%$ corresponds to $\Delta
Z_{\rm eff}=0.75$ and $M_{\imp}=0.63$).  Accounting for another
impurities present in the plasma leads to a small further decrease of
$M_{\imp}$, $M_{\imp} \approx 0.75$.  Thus, for similar experimental
conditions the transition from MH to MC is expected to occur at
somewhat higher $\He$ concentrations for the new JET wall in
comparison with those reported for the old carbon wall.  Along with
the approximate formula~(\ref{eq:x2.in.3heh}),
Eq.~(\ref{eq:Mimp.imps}) can serve as a rough estimate for the
transition concentration of helium-3 ions in $\HeH$ plasmas for
JET-like plasmas.

%\begin{figure}[htbp]
%\centering
%\includegraphics[trim=0.0cm 0.0cm 0.0cm 0cm, clip=true, scale=0.8]{figs/species.eps}
%\caption{$\zn_{i}=Z_i/A_i$ ratio for different ion species.}
%\end{figure}

\section{Suggestion for puffing ${\bf ^4 He}$ ions to $\HeH$ plasma
to control transition from MH to MC regime}
\label{sect:proposal}

In~view of the temperature dependence of $\Xcrhe$ shown in
Fig.~\ref{fig:xcrit.vs.param}(b) and the arguments concerning the
price of $\Xhe$, using smaller helium-3 concentrations for running
$\HeH$ experiments is beneficial. Depending on the specific goals
envisaged for the ICRH system one might need to set the experimental
conditions to reach either MH ($X[\He] < \Xcrhe$) or MC heating
($X[\He] > \Xcrhe$): operating in MH regime is preferred to increase
the fraction of thermal ion heating in $\HeH$ plasmas, whereas MC
could be potentially used for driving the current or local plasma
control.  Therefore, the development of the methods to decrease and
possibly control the minority concentrations needed for efficient ICRH
performance is of  high importance. As discussed in the previous
section of the paper, plasma dilution with impurities leads to the
reduction of $X[\He]$, at which the transition from minority heating
to mode conversion occurs. This section discusses briefly the
possibility of $\Xcrhe$ reduction by puffing additional gas to $\HeH$
mixture.

The concentration of intrinsic $\Be$ impurities in a plasma is
determined mostly by the level of plasma-wall interaction.  It cannot
be changed in a wide range and within a short time period that is a
prerequisite to control the transition concentration of minority
ions. However, additional puffing of $^{4}{\rm He}$ ions to $\HeH$
plasma seems to potentially fulfil this goal.  $^{4}{\rm He}$ is a
D-like species, and its price is much lower than the price for
helium-3.  As follows from Eq.~(\ref{eq:Mimp.imps}), we expect that
$\Xcrhe$ decreases with the concentration of helium-4 ions according
to \beq M_{\imp} \approx 1 - 8X[\Be] - 4.9 X[^{4}{\rm He}].
\label{eq:Mimps.4he}
\eeq

\begin{figure}[htbp]
\centering
\includegraphics[trim=0.0cm 0.0cm 0cm 0.0cm, clip=true, width=10cm]{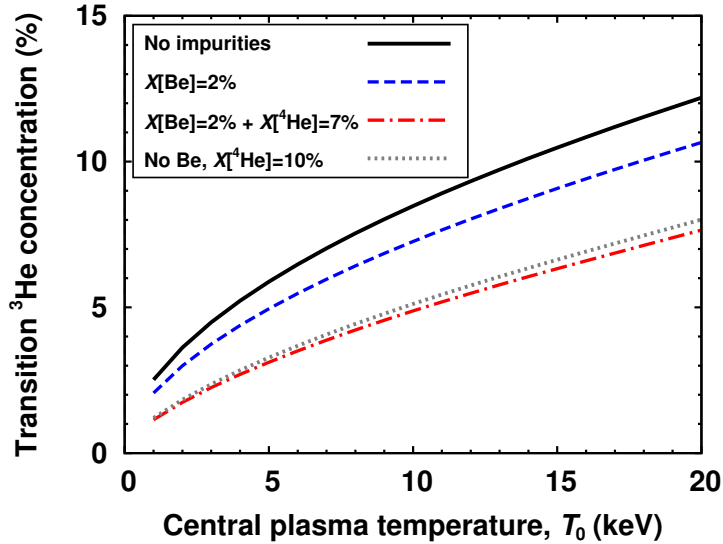}
\caption{Transition concentration of helium-3 ions in $\HeH$ plasma
increases with raising the plasma temperature. By puffing $^{4}{\rm He}$
ions, $X_{\crit}[\He]$ can potentially be reduced and/or controlled.
\label{fig:proposal}
}
\end{figure}

Figure~\ref{fig:xcrit.vs.c.4he}(b) shows the comparison of the results
given by Eq.~(\ref{eq:Mimps.4he}) and the values calculated
numerically with the TOMCAT solver for the baseline parameters
considered ($f=32.2\,{\rm MHz}$ and $B_0=3.1\,{\rm T}$).  In plasmas
without Be ions, addition of 5\% and 10\% of helium-4 ions to the
$\HeH$ plasma decreases the transition concentration $\Xcrhe$ by 22\%
and 44\%, respectively.  Accounting for the presence of Be ions in the
plasma, even smaller fractions of helium-4 can be used for minority
ion control. As Fig.~\ref{fig:proposal} illustrates, for plasmas
including $\Be$ impurities at the level of 2\%, by puffing 7\% of
$^{4}{\rm He}$ one obtains similar transition concentration $\Xcrhe$
as those in pure $\HeH$ plasmas with $X[^{4}{\rm He}]=10\%$ puffing.

If a fraction of fuel hydrogen ions are replaced by $^{4}{\rm He}$
ions the effective charge $Z_{\rm eff}$ will raise. But at the same
time the level of $\He$, which also contributes to $Z_{\rm eff}$,
decreases, so the resulting increase of the effective charge is
reasonably small, $\Delta Z_{\rm eff} \approx 0.1-0.15$.  Therefore it seems to be worth exploring this option in detail in separate studies.

A similar potential method to control $\Xcrhe$ could be applied if operating with
deuterium majority plasmas, in which helium-3 is partially burn down
via the fusion reaction
${\rm D} + {\rm ^3 He} \rightarrow {\rm ^4 He}(3.6\,{\rm MeV}) + p (14.7\,{\rm MeV})$.
%and
%${\rm D} + {\rm ^3 He} \rightarrow {\rm ^5 Li} + \gamma(16.4\,{\rm MeV})$.
By~adding $X[\Hy]=15\%$ to $\HeD$ plasmas, the transition concentration of helium-3 ions
in such plasmas can be reduced by a factor of $\sim 0.75-0.8$.

\section{Conclusions}
\label{sect:concls}

ITER will start its operation using the hydrogen majority plasmas to
minimize the activation of the tokamak components.  Since the access
to H-mode in that plasmas will not be assured with the available
heating powers, it is of high importance to maximize the heating
performance for the non-activated phase of ITER operation.  ICRH is
one of the heating systems to be used in ITER, and a number of ICRH
experiments were performed at JET to develop and optimize heating
scenarios relevant for such plasmas.  One of the promising ICRH
schemes relies on the use of the resonant minority $\He$ ions to
absorb the RF energy.

Past JET experiments with the old wall consisting of carbon tiles
highlighted a~number of specific features when applying ICRH in $\HeH$
plasmas. Particularly, a~significant effect of intrinsic carbon
impurities was revealed in that experiments.  The heating region was
found to be shifted radially comparing to the heating maximum expected
in a pure plasma, and that was attributed to the presence of
impurities.  Under certain conditions C impurities also produced a
supplementary MC layer, and multiple MC dynamics was observed in
$\HeH$ plasmas. In addition, the concentration of minority $\He$ ions,
at which the transition from MH to MC occurs, was reported to be lower
than that predicted by numerical simulations if the effect of
impurities was neglected.

In the present paper we discuss how the transition concentration of
$\He$ ions depends on the plasma and ICRH parameters. It is shown that
$\Xcrhe$ is related to the Doppler broadening of the minority IC
layer, and thus is inversely proportional to the antenna frequency and
increases with the plasma temperature and FW toroidal wavenumber.
Using two equivalent analytical approaches, we generalize the formula
for the transition minority concentration given  in
Ref.~\cite{wesson} for the case of inverted ICRH scenarios, which
$\HeH$ scheme belongs to.  This allows us to explain a small increase
of $\Xcrhe$ with the plasma density in $\HeH$ plasmas.

Accounting for multiple impurity species always present in the plasma,
we show that $\Xcrhe$ decreases and scales almost linearly with the
impurity concentrations. An analytical estimate for the relative
change of $\Xcrhe$ due to impurities is derived, and is shown to be in
a good correspondence with the numerical results. We demonstrate that
$\Be$ is to be the main impurity species affecting $\Xcrhe$ in $\HeH$
plasmas for JET equipped with the new ITER-like wall.  A reduction of
$\Xcrhe$ by $\sim20-25\%$ is predicted if considering typical Be and
another impurity concentrations at JET. A~possible method to reduce
and control $\He$ level, at which the transition from MH to MC is to
occur, is suggested: the method relies on the additional puffing of
$^{4}{\rm He}$ ions to $\HeH$ plasmas.  We show that for $X[^{4}{\rm
  He}]=5-10\%$ the reduction of the transition concentration of helium-3
ions by a factor of $\sim 0.6$ can be expected.

Since a~number of simplifying assumptions have been made in this
paper, the results can serve only as a qualitative estimates rather
quantitative predictions.  However, our results support the earlier
experimental findings and allows to understand the basics of the
underlined ICRH physics and interpolate the discussed effect
accounting for new impurity species present in a~plasma.  Further
numerical modelling and experimental support is needed to confirm the
potential of using $^{4}{\rm He}$ gas puffing for $\Xcrhe$ control in
$\HeH$ plasmas.

\section{Acknowledgements}
\label{sect:acknowls}
This work was funded by the European Communities under Association Contract
between EURATOM and \emph{Vetenskapsr{\aa}det}. The views and opinions
expressed herein do not necessarily reflect those of the European Commission.

\section*{Appendix A. IIH and L-cutoff frequencies accounting
for multiple impurity species and ${\bf n_{tor} \neq 0}$}
\label{sect:appA}
\appendix
\setcounter{section}{1}

In the IC frequency range the cold plasma tensor components are given by
\beq
\eqalign{
\epsS= 1 + \frac{\ompe^2}{\omce^2} - \sum_{i} \frac{\ompi^2}{\om^2 - \omci^2}, \\
\epsL= 1 + \frac{\ompe^2}{\omce^2} - \sum_{i} \frac{\ompi^2}{\omci(\om - \omci)}.
\label{eq:A1}
} \eeq The sum in~(\ref{eq:A1}) is to be taken over all ion species,
including impurities.  The conditions $\epsS = \npar^2$ and $\epsL =
\npar^2$ defining the location of the IIH resonance and L-cutoff
can be re-written in a simpler form \beq \sum_{i} \frac{f_i
  \zn_i}{\zn_i^2 - \omn_{\rm S}^2} = \alpha, \\ \sum_{i}
\frac{f_i}{\zn_i - \omn_{\rm L}} = \alpha,
\label{eq:A2}
\eeq
by introducing the following notations for the ion species `$i$':
$X_{i}=n_{i}/n_{e}$, $f_{i} = Z_{i} X_{i}$ -- the fraction
of the replaced electrons,
$\zn_{i}=Z_{i}/A_{i}$ -- ratio of the charge number to the
atomic mass;
$\omc{\rm H} = eB/(m_{\rm H}c)$ -- the cyclotron frequency of hydrogen ions that
is used for the frequency normalization ($\omn = \om/\omc{\rm H}$),
$\omp{\rm H} = \sqrt{4\pi n_{e} e^2/m_{\rm H}}$ --
the reference hydrogen plasma frequency (note $n_e$ instead of $n_{\rm H}$),
$\tilde{n}_{\|}^2 = n_{\|}^2\,-\,1\,- \ompe^2/\omce^2$  --
effective $n_{\|}^2$ if accounting for the small vacuum and electron contributions
to the tensor component $\epsS$. Then, the small parameter
appearing on the right-hand side of Eqs.~(\ref{eq:A2}), which
describes the effect of finite $\ntor$ on the location
of the IIH resonance and L-cutoff, is given by
$\alpha = (\omc{\rm H}^2 / \omp{\rm H}^2) \,{\tilde{n}_{\|}^2}$.

By substituting $f_1 = 1 - f_2 - \sum\limits_{\imp} f_{\imp}$,  it can be
shown that Eqs.~(\ref{eq:A2}) are equivalent to the following ones
\beq
\eqalign{
-\zn_1 + \frac{(\zn_1 - \zn_2)(\omns^2 + \zn_1 \zn_2)}{\omns^2 - \zn_2^2} f_2 = \gamma_{\rm S}, \\
\gamma_{\rm S} = \alpha(\omns^2 - \zn_1^2) - \sum_{\imp} \frac{(\zn_1 - \zn_{\imp})(\omns^2 + \zn_1 \zn_{\imp})}{\omns^2 - \zn_{\imp}^2} f_{\imp}
\label{eq:A3}
}
\eeq
and
\beq
\eqalign{
-1 + \frac{\zn_1 - \zn_2}{\omnl - \zn_2}\,f_2 = \gamma_{\rm L}, \\
\gamma_{\rm L} = \alpha(\omnl - \zn_1) - \sum_{\imp} \frac{\zn_1 - \zn_{\imp}}
{\omnl - \zn_{\imp}}\,f_{\imp}.
\label{eq:A4}
} \eeq The right-hand side of Eqs.~(\ref{eq:A3}) and (\ref{eq:A4})
includes contributions due to finite $\ntor$ and
impurities. Neglecting those terms the well-known expressions for the
IIH and L-cutoff frequencies are obtained \beq \eqalign{ \wns = \zn_2 \sqrt{\frac{1
      - (1 - \zn_1/\zn_2)f_2}{1 - (1 - \zn_2/\zn_1)f_2}}
  \approx \\
  \quad \quad \approx \zn_2 + \frac{(\zn_1^2 - \zn_2^2)}{2\zn_1} f_2 -
  \frac{(\zn_1 + \zn_2)(\zn_1 - \zn_2)^2(\zn_1 - 3\zn_2)}{8\zn_1^2 \zn_2} f_2^2, \\
  \wnl = \zn_2 + (\zn_1 - \zn_2) f_2.
\label{eq:A.wnsl.clas}
}
\eeq

For a plasma consisting of $N$ ion species with different $Z/A$ ratios
and if $\ntor \neq 0$, there are $N$ different solutions for $\omns^2$
and $\omnl$ satisfying \,Eqs.~(\ref{eq:A3}) and (\ref{eq:A4}). Thus,
in the general case the solutions has to be evaluated numerically.
There is an artificial solution, which appears due to $\alpha \neq 0$
$(\omnl \simeq -1/\alpha,$\,\, $\omns^2 \simeq -\zn_1/\alpha)$, while
the other $N-1$ solutions represent IIH resonances and L-cutoffs
associated with the minority ions and $N-2$ impurity species.  We are
interested in the solutions for $\omns$ and $\omnl$, which correspond
to the minority ions and generalize Eqs.~(\ref{eq:A.wnsl.clas}),
aiming for identifying the influence of $\ntor$ and impurities on
that.

Omitting the comprehensive algebra, we provide the formula
for L-cutoff valid up to quadratic terms in $\alpha$ and $f_{\imp}$,
including the cross-terms:
\beq
\eqalign{
\omn_{\rm L} = \wnl + \eps_{\rm L}, \\
\eps_{\rm L} = \lpb k_{11} \alpha f_2 (1- f_2) + \sum_{\imp} k_{12} f_2 f_{\imp} \rpb
\times \lpb 1 + k_{21} \alpha  + \sum_{\imp} k_{22} f_{\imp} \rpb,
\label{eq:A.wnl.quadr}
}
\eeq
where the parameters $k_{ij}$ are given by
\beq
\eqalign{
k_{11} = (\zn_1 - \zn_2)^2, \\
k_{12} = (\zn_1 - \zn_2)(\zn_1 - \zn_{\imp})/(\wnl - \zn_{\imp}), \\
k_{21} = (\zn_1 - \zn_2)(1-2f_2), \\
k_{22} = (\zn_1 - \zn_{\imp})(\zn_2 - \zn_{\imp})/(\wnl - \zn_{\imp})^2.
\label{eq:A.kij.for.wl}
}
\eeq
%
%
%\beq
%\eqalign{
%\omn_{\rm L} = \wnl + \lpb \alpha (\zn_1 - \zn_2)^2 f_2 (1-f_2) +
%\sum\limits_{\imp} \frac{(\zn_1 - \zn_2)(\zn_1 - \zn_{\imp})}{(\wnl - \zn_{\imp})}f_2 f_{\imp} \rpb \times \\
%\times \lpb 1 + \alpha (\zn_1 - \zn_2)(1 - 2f_2) + \sum\limits_{\imp} \frac{(\zn_1 - \zn_{\imp})(\zn_2 - %\zn_{\imp})}{(\wnl - \zn_{\imp})^2} \fimp \rpb
%.}
%\eeq
%
The expression for the IIH resonance is more complicated since
Eq.~(\ref{eq:A3}) is written for $\omns^2$ rather than $\omns$.
In that case, it can be shown that
\beq
\eqalign{
\omns = \sqrt{\wns^2 + \eps_{\rm S}}\,, \\
\eps_{\rm S} = \lpb k_{11} \alpha f_2 (1- f_2) + \sum_{\imp} k_{12} f_2 f_{\imp} \rpb
\times \lpb 1 + k_{21} \alpha  + \sum_{\imp} k_{22} f_{\imp} \rpb
%\eps_{\rm S} = \alpha \zn_1 \zn_2 (\zn_1^2 - \zn_2^2)^2 f_2 (1-f_2)/[...]^3
%+ \sum_{\imp} \frac{\zn_2(\zn_1^2 - \zn_2^2)(\zn_1 - \zn_{\imp})
%(\wns^2 + \zn_1 \zn_{\imp})}{(\wns^2 - \zn_{\imp}^2)} f_2 f_{\imp}
\label{eq:A.wns.quadr}
}
\eeq
is a high-accuracy quadratic approximation for
 $\omns$.
The expansion coefficients $k_{ij}$ for $\omns$ are given by
\beq
\eqalign{
k_{11}\,\bar{\zn}_1^3 = \zn_1 \zn_2 (\zn_1^2 - \zn_2^2)^2,  \\
k_{12}\,\bar{\zn}_1^2 = \zn_2 (\zn_1^2 - \zn_2^2)(\zn_1 - \zn_{\imp})(\wns^2 + \zn_1 \zn_{\imp})
/(\wns^2 - \zn_{\imp}^2),  \\
k_{21}\,\bar{\zn}_1^2 = (\zn_1^2 - \zn_2^2)(\zn_1 - (\zn_1 + \zn_2)f_2),  \\
k_{22}\,\bar{\zn}_1 = (\zn_1 - \zn_{\imp})/(\wns^2 - \zn_{\imp}^2)^2\, \cdot \,
\left[ (\wns^2 - \zn_{\imp}^2)(\wns^2 + \zn_1 \zn_{\imp}) - \right. \\
\quad \left. - \zn_{\imp}(\zn_1 + \zn_{\imp})(\wns^2 - \zn_2^2) \right],
\label{eq:A.kij.for.ws}
}
\eeq
%$$
%k_{22}\,[...] = (\zn_1 - \zn_{\imp})
%\lpb \wns^4 - 2\zn_{\imp}^2 \wns^2 + \zn_2^2 \zn_{\imp}^2 + \zn_{1}\zn_{\imp}(\zn_2^2 - \zn_{\imp}^2) \rpb
%/(\wns^2 - \zn_{\imp}^2)^2,
%$$
%
where $\bar{\zn}_1 = \zn_1 - (\zn_1 - \zn_2)f_2$.
Note that Eqs.~(\ref{eq:A.wnl.quadr}--\ref{eq:A.kij.for.ws})
are valid for an arbitrary number~of impurity species, and
include the impurity and $\ntor$ contributions simultaneously.

\begin{figure}[htbp]
\centering
\includegraphics[trim=0.0cm 0.0cm 0cm 0.0cm, clip=true, width=7.5cm]{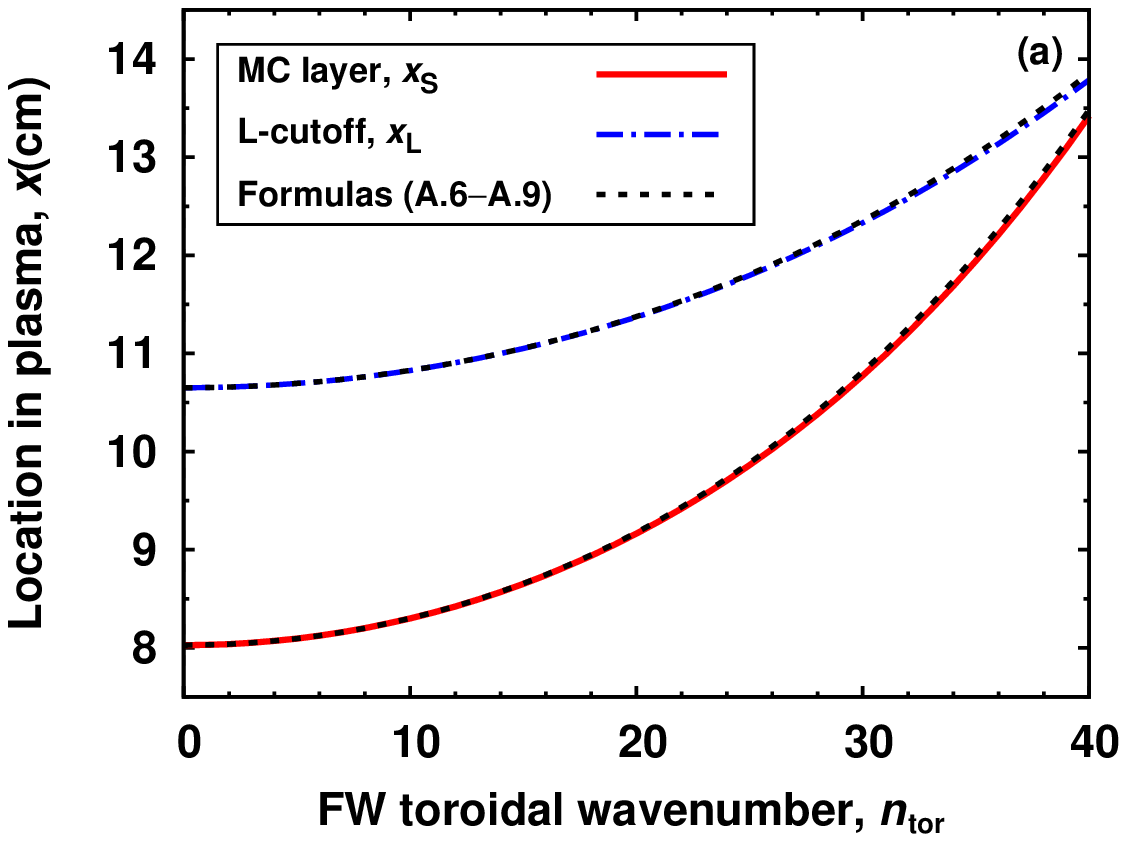}
\includegraphics[trim=0.0cm 0.0cm 0cm 0.0cm, clip=true, width=7.5cm]{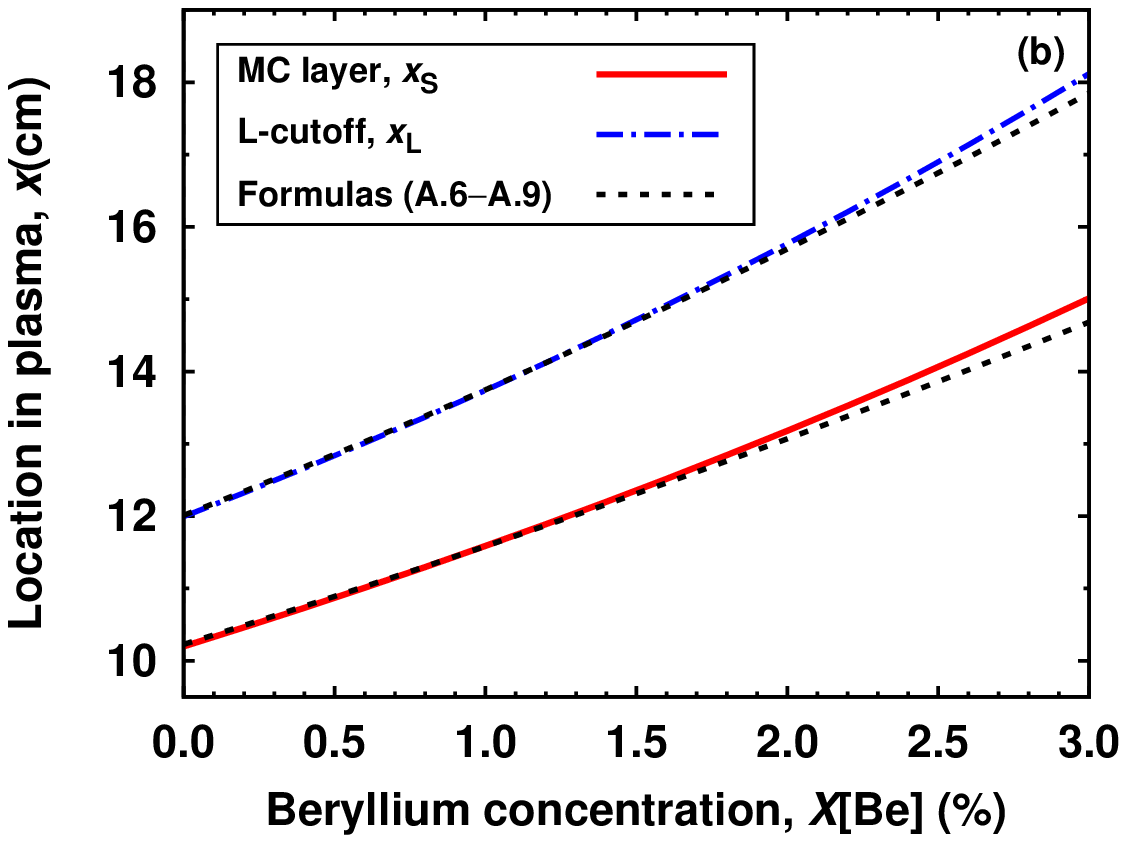}
\caption{Location of the IIH resonance and L-cutoff in $\HeH$ plasma\,\,
($X[\He]~=~5.9\%$) as a function of the FW toroidal wavenumber (a), and
beryllium concentration ($\ntor=27$) (b).
\label{fig:appA}
}
\end{figure}

Figure~\ref{fig:appA}(a) shows the location of the IIH resonance and
L-cutoff in pure $\HeH$ plasma as a function of the FW toroidal
wavenumber (the other parameters correspond to those used in
Fig.~\ref{fig:pabs.vs.x3he}).  Considering $\kpar \neq 0$, both the
IIH resonance and L-cutoff shift towards the LFS, and the thickness of
the evanescence layer gradually decreases.  As Fig.~\ref{fig:appA}(a)
clearly illustrates, this shift is almost linear in $\alpha$ and thus
quadratic in $\ntor$.  Analytical approximations given by
Eqs.~(\ref{eq:A.wnl.quadr}--\ref{eq:A.kij.for.ws}) are shown in
Fig.~\ref{fig:appA} with dots and are in excellent agreement with the
numerical results for the whole range of experimentally relevant
$\ntor$. The formulas~(\ref{eq:A.wnl.quadr}--\ref{eq:A.kij.for.ws})
are in an almost perfect agreement with numerical results if also
impurities are accounted for. Figure~\ref{fig:appA}(b) is calculated
for $\ntor=27$ and illustrates further (almost linear) shift of the MC
layer and \mbox{L-cutoff} towards the LFS if the beryllium concentration is
increased in a $\HeH$ plasma. In~contrast to the effect of $\ntor$,
the width of the evanescence layer increases gradually with $X[\Be]$.

If $f_2\ll 1$ Eqs.~(\ref{eq:A.wns.quadr}) and (\ref{eq:A.kij.for.ws})
may be simplified by keeping in $\eps_{\rm S}$ only the dominant terms in
$f_2$: \beq \eqalign{ \omns^2 \approx \zn_2^2 + \frac{\zn_2(\zn_1^2 -
    \zn_2^2)}{\zn_1}f_2
  + \frac{\zn_2(\zn_1+\zn_2)(\zn_1 - \zn_2)^2}{\zn_1^2}f_2^2\,+\,\\
  \quad \quad +\, f_2 \lpb k_{11} \alpha + \sum_{\imp} k_{12} f_{\imp} \rpb \times \lpb 1 + k_{21} \alpha + \sum_{\imp} k_{22} f_{\imp} \rpb, \\
  k_{11} \approx \zn_2(\zn_1^2 - \zn_2^2)^2/\zn_1^2, \\
  k_{12} \approx \zn_2(\zn_1^2-\zn_2^2)(\zn_1 - \zn_{\imp})(\zn_2^2 + \zn_1 \zn_{\imp})/(\zn_1^2(\zn_2^2 - \zn_{\imp}^2)), \\
  k_{21} \approx (\zn_1^2 - \zn_2^2)/\zn_1, \\
  k_{22} = k_{12}k_{21}/k_{11} \approx (\zn_1 - \zn_{\imp})(\zn_2^2 +
  \zn_1 \zn_{\imp})/(\zn_1 (\zn_2^2-\zn_{\imp}^2)).
\label{eq:A.wns.quadr.f2}
}
\eeq
To the lowest order (linear in $\alpha$ and $f_{\rm imp}$)
we have
\beq
\eqalign{
\fl{
\omnl \approx \zn_2 + (\zn_1 - \zn_2)f_2 + (\zn_1 - \zn_2)^2\,\alpha f_2
+ \sum_{\imp} \frac{(\zn_1 - \zn_2)(\zn_1 - \zn_{\imp})}{\zn_2 - \zn_{\imp}} f_2 f_{\imp}\,,} \\
\fl{\omns \approx \zn_2 + \frac{\zn_1^2 - \zn_2^2}{2\zn_1}f_2 +
\frac{(\zn_1^2 - \zn_2^2)^2}{2\zn_1^2}\,\alpha f_2\,+ } \\
+ \sum_{\imp} \frac{(\zn_1^2 - \zn_2^2)(\zn_1 - \zn_{\imp})(\zn_2^2+\zn_1 \zn_{\imp})}
{2\zn_1^2 (\zn_2^2 - \zn_{\imp}^2)} f_2 f_{\imp}.
\label{eq:A.wnsl.linear}
} \eeq The first terms in Eqs.~(\ref{eq:A.wnsl.linear}) show that both
the IIH resonance and L-cutoff are located close to the minority IC
layer. The second term illustrates a successive shift of the MC and
L-cutoff layers towards the HFS or the LFS (depending on $\zn_1/\zn_2$
ratio) with increasing the concentration of minority ions.  The third
term corresponds to the contribution due to finite $\ntor$.  It is
always positive and, thus, accounting for $\kpar \neq 0$ results in
a small shift of the layers towards the LFS.  Finally, the fourth
term represents the effect of impurities on the location of the layers
in a plasma; its sign, which defines the direction of the additional
shift due to impurities, depends on the sign of the ratio $(\zn_1 -
\zn_2)(\zn_1 - \zn_{\imp})/(\zn_2 - \zn_{\imp})$.
Equation~(\ref{eq:A.wnsl.linear}) for $\omns$ is used in the main text
of the paper to evaluate the transition concentration of helium-3 ions
in $\HeH$ plasma and the effect of impurities on that.

\vspace{-0.3cm}
\section*{Appendix B. On `$\pm$' sign in formulas
for the transition minority concentration}
\label{sect:appB}
\appendix
\setcounter{section}{2}

\begin{figure}[htbp]
\centering
\includegraphics[trim=0.0cm 0.0cm 0cm 0.0cm, clip=true, width=12.0cm]{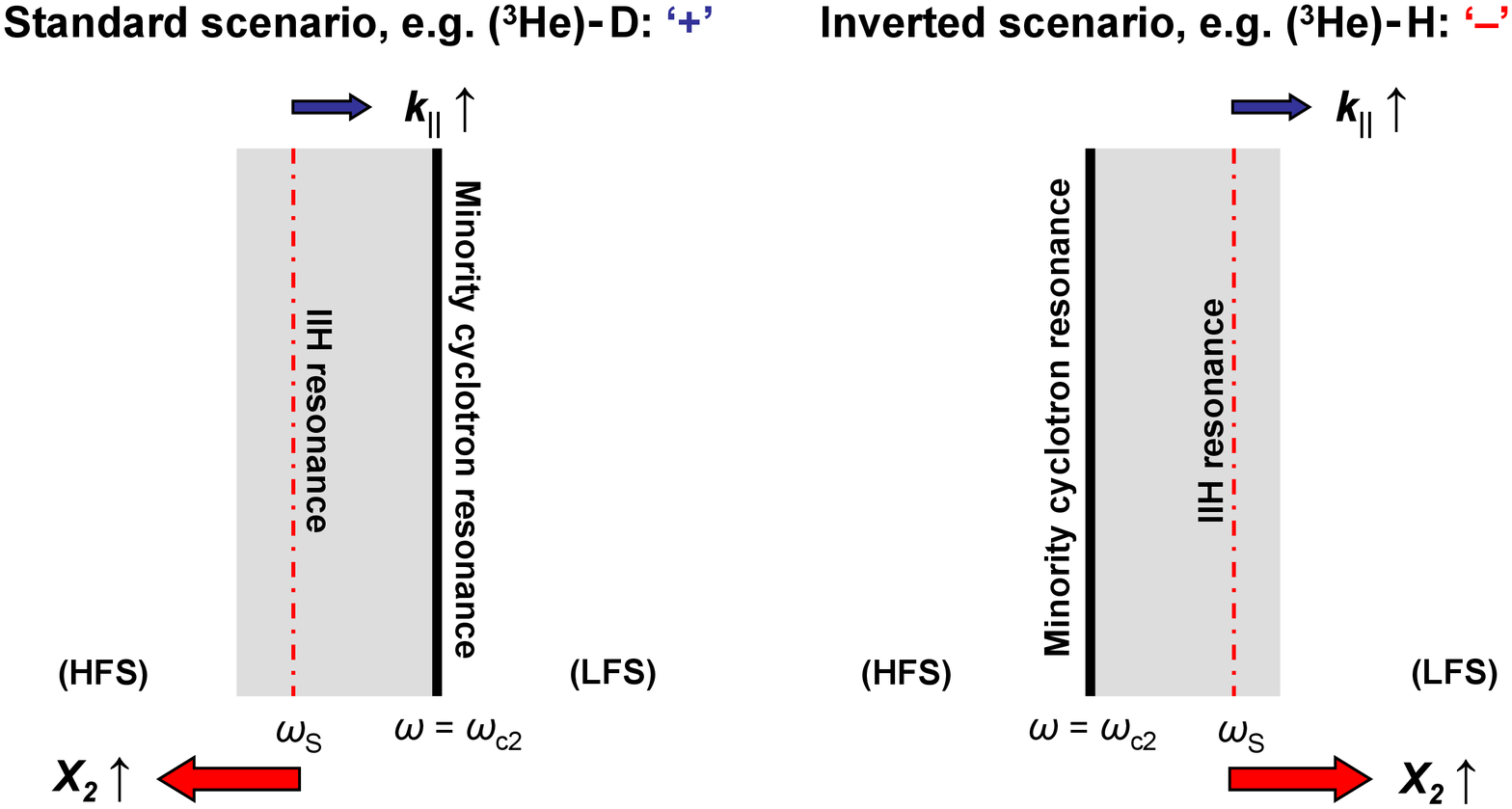}
\caption{The shift of the IIH resonance with increasing $X_2$ and $\kpar$
is in the opposite/same direction for the standard/inverted
ICRH scenarios, respectively. It~results in a `$\pm$'
sign that appears in the correction term for the
transition minority concentration given by Eqs.~(\ref{eq:x2.wesson}) and (\ref{eq:x2.kazakov.general}).
\label{fig:appB}
}
\end{figure}

There is a simple physical explanation for the appearance of `$\pm$'
sign in Eqs.~(\ref{eq:x2.wesson}) and (\ref{eq:x2.kazakov.general})
for the transition minority concentration. As mentioned in
section~\ref{sect:xcrit.theory}, `$+$' or `$-$' sign is to be taken
for the standard and inverted ICRH scenarios, respectively.  In
plasmas without impurities the IIH frequency is given by \beq
\omn_{\rm S} \approx \zn_2 + \frac{\zn_1^2 -\zn_2^2}{2\zn_1} f_2 +
\frac{(\zn_1^2 - \zn_2^2)^2}{2\zn_1^2}\,\alpha f_2,
\label{eq:B1}
\eeq where $\alpha \propto (\ntor^2/n_{e})(B/f)^2$.  The sign of the
second term in Eq.~(\ref{eq:B1}) is different for the standard ($\zn_1
< \zn_2$) and inverted ($\zn_1 > \zn_2$) ICRH scenarios.  While for
the standard scenarios the MC layer shifts towards the HFS with
increasing the minority concentration, for the inverted scenarios it
moves in the opposite (LFS) direction (Figure~\ref{fig:appB}). At the
same time the shift of the MC layer due to finite $k_{\|}$ (described
by the third term in Eq.~(\ref{eq:B1})) is always towards the LFS
regardless the relation of $Z/A$ ratio for majority and minority ions.
Thus, for the inverted ICRH scenarios the shift of the MC layer due to
$X_2$ and $\kpar$ is in the same direction, and therefore smaller
minority concentration is needed for the MC layer to pass through the
Doppler broadened minority IC region. Vice versa, for the standard
scenarios the $k_{\|}$ correction counteracts the HFS shift of the MC
layer with increasing $X_2$. This results in higher minority
concentrations needed to pass the border, which marks the transition
from~MH~to~MC.

%%%%%%%%%%%%%%%%%%%%%%%%%%%%%%%%%%%%%%%%%%%%%%%%%%
%%%%%%	References:
%%%%%%%%%%%%%%%%%%%%%%%%%%%%%%%%%%%%%%%%%%%%%%%%%%

\end{document}